\documentclass[12pt]{article}
\usepackage{amssymb}
\usepackage{amsmath}
\usepackage{stmaryrd}
\usepackage{bbm}
\usepackage{graphicx}
\catcode`@=11 
\renewcommand{\section}{\@startsection{section}{1}{0pt}{\medskipamount}
{\medskipamount}{\bf}}
\numberwithin{equation}{section}
\catcode`@=12 
\def\a{\alpha}
\def\b{\beta}

\def\de{\delta}
\def\e{\epsilon}
\def\f{\phi}
\def\g{\gamma}
\def\p{\psi}

\def\la{\lambda}

\def\s{\sigma}

\newcommand{\C}{\mathbb C}
\newcommand{\R}{\mathbb R}

\newcommand{\cN}{{\cal N}}
\newcommand{\unity}{\mathbbm{1}}
\def\be{\begin{equation}}
\def\ee{\end{equation}}
\def\arr{\begin{array}{rll}}
\def\ea{\end{array}}
\def\bea{\begin{eqnarray}}
\def\eea{\end{eqnarray}}
\def\sfrac#1#2{{\textstyle\frac{#1}{#2}}}

\def\ph{\phantom{-}}
\def\ic{{\rm i}}

\def\pa{\partial}
\def\>{\rangle}
\def\<{\langle}
\def\+{\dagger}
\def\={\ =\ }
\def\und{\qquad\textrm{and}\qquad}
\def\ax{\a{\cdot}x}
\def\bx{\b{\cdot}x}
\def\gx{\g{\cdot}x}
\def\uh{U_{\rm{hom}}}

\def\bt{\bar{\tau}}
\def\hx{\hat{x}}
\def\1d{{\dot{1}}}
\def\2d{{\dot{2}}}
\begin{document}
\renewcommand{\thefootnote}{\fnsymbol{footnote}}
\begin{titlepage}
\setcounter{page}{0}
\begin{flushright}
CERN-PH-TH/2008-041\\
ITP--UH--04/08\\
LMP-TPU--01/08\\
\end{flushright}
\vskip 1cm
\begin{center}
{\LARGE\bf N=4 mechanics, WDVV equations and roots}
\vskip 2cm
$
\textrm{\Large Anton Galajinsky\ }^{a} ,\quad
\textrm{\Large Olaf Lechtenfeld\ }^{b\ c} ,\quad
\textrm{\Large Kirill Polovnikov\ }^{a}
$
\vskip 0.7cm
${}^{a}$ {\it
Laboratory of Mathematical Physics, Tomsk Polytechnic University, \\
634050 Tomsk, Lenin Ave. 30, Russian Federation} \\
{Emails: galajin, kir @mph.phtd.tpu.edu.ru}
\vskip 0.4cm
${}^{b}$ {\it
Theory Division, Physics Department, CERN, \\
1211 Geneva 23, Switzerland} \\
{Email: olaf.lechtenfeld@cern.ch}
\vskip 0.4cm
${}^{c}$ {\it
Institut f\"ur Theoretische Physik, Leibniz Universit\"at Hannover,\\
Appelstrasse 2, D-30167 Hannover, Germany} \\
{Email: lechtenf@itp.uni-hannover.de}
\vskip 0.2cm
\end{center}
\vskip 2cm
\begin{abstract} \noindent
$\cN{=}4$ superconformal multi-particle quantum mechanics on the real line
is governed by two prepotentials, $U$ and $F$, which obey a system of
partial differential equations linear in~$U$ and generalizing the 
Witten-Dijkgraaf-Verlinde-Verlinde (WDVV) equation for~$F$. 
Putting $U{\equiv}0$ yields a class of models (with zero central charge) 
which are encoded by the finite Coxeter root systems. 
We extend these WDVV~solutions~$F$ in two ways:
the $A_n$ system is deformed $n$-parametrically to the edge set
of a general orthocentric $n$-simplex, and the $BCF$-type systems
form one-parameter families. A classification strategy is proposed.
A nonzero central charge requires turning on~$U$ in a given $F$~background,
which we show is outside the reach of the standard root-system ansatz for 
indecomposable systems of more than three particles. In the three-body case, 
however, this ansatz can be generalized to establish a series of nontrivial 
models based on the dihedral groups~$I_2(p)$, which are permutation symmetric
if 3 divides~$p$. We explicitly present their full prepotentials. 
\end{abstract}
\vspace{2cm}
\end{titlepage}
\renewcommand{\thefootnote}{\arabic{footnote}}
\setcounter{footnote}0

\section{Introduction}
\noindent 
It has been known for a long time that integrable quantum systems are 
intimately related to Lie algebras (see, for instance, \cite{olpe}).
Therefore, it is natural to expect their appearance also in supersymmetric
extensions of integrable multi-particle quantum mechanics models. 
In this paper, we revisit such systems with $\cN{=}4$~{\it superconformal\/} 
symmetry in one space dimension and, within a canonical ansatz, investigate 
them for the superconformal algebra~$su(1,1|2)$ with central charge~$C$. 
Despite physical interest in these models~\cite{gibtow}, 
their explicit construction has remained an open problem until now.

$\cN{=}4$ superconformal many-body quantum systems on the real line are very 
rigid. Their existence is governed by a system of nonlinear partial 
differential equations for two prepotentials, $U$ and~$F$, for which 
few solutions are known when $C{\neq}0$~\cite{wyl,bgk,bgl,glp2}. 
The determination of~$F$ is decoupled from~$U$ and requires solving `only' the 
well-known (generalized) Witten-Dijkgraaf-Verlinde-Verlinde (WDVV) 
equation~\cite{w,dvv}, which arises in topological and Seiberg-Witten 
field theory. The WDVV~solutions known so far are all based -- again -- 
on the root systems of simple Lie algebras~\cite{magra,vese}.\footnote{
As a slight generalization, all Coxeter reflection groups appear.}

If $C{=}0$ any WDVV solution~$F$, together with $U{\equiv}0$, will provide
a valid multi-particle quantum model. For nonzero central charge, however,
one is to solve a second partial differential equation for~$U$ in the
presence of~$F$. To this so-called `flatness conditon' only particular
solutions for at most four particles are in the literature~\cite{wyl,glp2}.

All considerations up to now have employed a natural ansatz for $F$ and~$U$
in terms of a set $\{\a\}$ of covectors. We find, however, that for
systems of less than four particles this ansatz must be generalized
in order to capture all solutions. In these cases, the WDVV equation is
trivially satisfied, and we can (and do) construct new three-body models for 
any dihedral $I_2(p)$~root system, starting with a Calogero-type $A_2$~model.
For more than three particles, where the WDVV equation is effective,
we show that even our generalized ansatz is insufficient to produce
irreducible $U{\neq}0$ solutions in the root-system context.
A model is reducible if, after removing the 
center-of-mass degree of freedom, it can be decomposed into decoupled
subsystems. As for the WDVV~equation alone, we generalize the 
solutions of~\cite{magra,vese} and give a geometric interpretation of 
certain $A_n$ deformations~\cite{chaves} in terms of orthocentric simplices.

The paper is organized as follows.
In Section~2 we recall the formulation of conformal mechanics of 
$n{+}1$~identical particles on the real line in terms of $so(1,2)$ generators
including the Hamiltonian. In this description, an $\cN{=}4$~supersymmetric
extension with central charge~$C$ is straightforward to construct as we 
demonstrate in Section~3. The closure of the superconformal algebra poses 
constraints on the interaction, which in Section~4 lead to what we call the 
`structure equations' on the prepotentials $U$ and~$F$. The analysis of these 
structure equations in Section~5 suggests constructing the prepotentials in 
terms of a system of covectors, which reduces the differential equations to 
nonlinear algebraic equations. Sections 6 and~7 derive families of 
$F$~solutions with $U{\equiv}0$, based on certain deformations of the root 
systems of the finite reflection groups. Turning on~$U$ for these 
$F$~backgrounds is analyzed in Sections 8 and~9, with negative results 
for more than three particles, but with a positive classification and the 
full construction of the prepotentials for three particles via the dihedral 
groups~$I_2(p)$, including five explicit examples. 
Section~10 concludes.
\vspace{1cm}

\section{Conformal quantum mechanics}
\noindent
Let us consider a system of $n{+}1$ identical particles with unit mass, 
moving on the real line according to a Hamiltonian of the generic form
($I=1,\ldots,n{+}1$)
\be\label{h}
H\=\sfrac{1}{2} p_I p_I\ +\ V_B (x^1, \dots, x^{n+1})\ .
\ee
Throughout the paper a summation over repeated indices is understood.
After separating the center-of-mass motion we will work with the $n$~degrees
of freedom of relative particle motion in later sections.
Also, the bosonic potential $V_B$ will get supersymmetrically extended to
a potential~$V$ including $V_B$.

For conformally invariant models the Hamiltonian~$H$ is a part of
the $so(1,2)$ conformal algebra 
\be\label{al} 
[D,H]\=-\ic H\ ,\quad 
[H,K]\=2\ic D\ ,\quad 
[D,K]\=\ic K\ , 
\ee 
where $D$ and $K$ are the dilatation and conformal boost generators,
respectively. Their realization in term of coordinates and
momenta, subject to 
\be 
[x^I, p_J]\=\ic  {\de_J}^I \ , 
\ee
reads 
\be 
D\=-\sfrac{1}{4} (x^I p_I +p_I x^I) \und 
K\= \sfrac{1}{2} x^I x^I \ . 
\ee 
The first relation in (\ref{al}) restricts the potential via \be\label{ucl}
(x^I \partial_I +2)\,V_B \=0\ , \ee meaning that $V_B$ must be
homogeneous of degree $-2$ for the model to be conformally
invariant. Imposing translation and permutation invariance and allowing
only two-body interactions, we arrive
at the Calogero model of $n{+}1$~particles interacting through 
an inverse-square pair potential,
\be 
V_B \= \sum_{I<J} \sfrac{g^2}{(x^I-x^J)^2} \qquad\longrightarrow\qquad 
H \= H_0 \ +\ V_B\ . 
\ee
\vspace{1cm}

\section{\cN=4 superconformal extension}
\noindent
Let us extend the bosonic conformal mechanics of the previous section
to an $\cN{=}4$ superconformal one,~\footnote{
For a one-particle model, see \cite{ikl}.}
with a single central extension~\cite{ioko}. 
The bosonic sector of the $\cN{=}4$ superconformal algebra $su(1,1|2)$
includes two subalgebras. Along with $so(1,2)$ considered in the previous 
section one also finds the $su(2)$ R-symmetry subalgebra generated by 
$J_a$ with $a=1,2,3$. The fermionic sector is exhausted by the SU(2) doublet 
supersymmetry generators $Q_\a$ and ${\bar Q}^\a$ as well as their 
superconformal partners $S_\a$ and ${\bar S}^\a$, with $\a=1,2$,
subject to the hermiticity relations 
\be
{(Q_\a)}^{\dagger}\={\bar Q}^\a \und {(S_\a)}^{\dagger}\={\bar S}^\a\ .
\ee
The bosonic generators are hermitian. The non-vanishing (anti)commutation 
relations in our superconformal algebra read\footnote{
$\s_1$, $\s_2$ and $\s_3$ denote the Pauli matrices.}
\begin{align}\label{algebra}
&
[D,H] \= -\ic \, H\ , && 
[H,K] \= 2\ic \, D\ ,
\nonumber\\[4pt]
&
[D,K] \= +\ic \, K\ , && 
[J_a,J_b] \= \ic \, \epsilon_{abc} J_c\ ,
\nonumber\\[2pt]
&
\{ Q_\a, \bar Q^\b \} \= 2\, H {\de_\a}^\b\ , &&
\{ Q_\a, \bar S^\b \} \=
+2\ic\,{{(\s_a)}_\a}^\b J_a-2\,D{\de_\a}^\b-\ic\,C{\de_\a}^\b\ ,
\nonumber\\[2pt]
&
\{ S_\a\,,\, \bar S^\b \} \= 2\, K {\de_\a}^\b\ , &&
\{ \bar Q^\a, S_\b \} \=
-2\ic\,{{(\s_a)}_\b}^\a J_a-2\,D{\de_\b}^\a+\ic\,C{\de_\b}^\a\ ,
\nonumber\\[2pt]
& 
[D,Q_\a] \= -\sfrac{1}{2} \ic\, Q_\a\ , && 
[D,S_\a] \= +\sfrac{1}{2} \ic \, S_\a\ ,
\nonumber\\[4pt]
&
[K,Q_\a] \= +\ic \, S_\a\ , && 
[H,S_\a] \= -\ic \, Q_\a\ ,
\nonumber\\[2pt]
&
[J_a,Q_\a] \= -\sfrac{1}{2} \, {{(\s_a)}_\a}^\b Q_\b\ , && 
[J_a,S_\a] \= -\sfrac{1}{2} \, {{(\s_a)}_\a}^\b S_\b\ ,
\nonumber\\[4pt]
& 
[D,\bar Q^\a] \= -\sfrac{1}{2} \ic \, \bar Q^\a\ , && 
[D,\bar S^\a] \= +\sfrac{1}{2} \ic \, \bar S^\a\ ,
\nonumber\\[4pt]
& 
[K,\bar Q^\a] \= +\ic \, \bar S^\a\ , && 
[H,\bar S^\a] \= -\ic \, \bar Q^\a\ ,
\nonumber\\[2pt]
&
[J_a,\bar Q^\a] \= \sfrac{1}{2} \, \bar Q^\b {{(\s_a)}_\b}^\a\ , && 
[J_a,\bar S^\a] \= \sfrac{1}{2} \, \bar S^\b {{(\s_a)}_\b}^\a\ .
\end{align}
Here $\e_{123}=1$, and $C$ stands for the central charge.

For a mechanical realization of the $su(1,1|2)$ superalgebra, 
one introduces fermionic degrees of freedom represented by the operators 
$\psi^I_\a$ and $\bar\psi^{I\a}$, with $I=1,\dots,n{+}1$ and $\a=1,2$, 
which are hermitian conjugates of each other and obey the anti-commutation 
relations\footnote{
Spinor indices are raised and lowered with the invariant
tensor $\e^{\a\b}$ and its inverse $\e_{\a\b}$, where $\e^{12}=1$.
}
\be
\{\p^I_\a, \p^J_\b \}\=0\ , \qquad 
\{ {\bar\p}^{I\a}, {\bar\p}^{J\b} \}\=0\ , \qquad
\{\p^I_\a, {\bar\p}^{J\b} \}\=\,{\de_\a}^\b \de^{IJ}\ .
\ee
In the extended space it is easy to construct the free fermionic generators 
associated with the free Hamiltonian $H_0=\frac{1}{2}p_Ip_I$, 
namely 
\be\label{QSfree}
{Q_0}_\a\=p_I \p^I_\a\ , \qquad 
\bar Q_0^\a\=p_I \bar\p^{I\a} \und
{S_0}_\a\=x^I \p^I_\a\ , \qquad 
\bar S_0^\a\=x^I \bar\p^{I\a}\ ,
\ee
as well as $su(2)$ generators
\be\label{Jfree}
{J_0}_a \= \sfrac{1}{2} \bar\p^{I\a} {{(\s_a)}_\a}^\b \p^I_\b\ .
\ee
Notice that these are automatically Weyl-ordered.
The free dilatation and conformal boost operators maintain their bosonic form
\be\label{DKfree}
D_0\= -\sfrac{1}{4}(x^I p_I +p_I x^I) \und K_0\= \sfrac12 x^I x^I \ .
\ee

In contrast to the $\cN{\le}2$ cases, the free generators fail to satisfy
the full algebra~(\ref{algebra}). Even for $C{=}0$, the $\{Q,\bar S\}$ and
$\{\bar Q,S\}$ anticommutators require corrections to the fermionic generators,
which are cubic in the fermions and can be restricted to $Q$ and~$\bar Q$ via
\be\label{Qcorr}
Q_\a \= Q_{0\a} -\ic\,[S_{0\a},V] \und
\bar Q^\a \= \bar Q^\a_0 - \ic\,[\bar S^\a_0,V] 
\qquad\textrm{where}\quad H \= H_0 + V
\ee
and $V\neq0$. Hence, there does not exist
a free mechanical representation of the algebra~(\ref{algebra}).
It follows further that $V$ contains terms quadratic and quartic in the
fermions, thus can be written as~\cite{wyl,bgl,glp2}\footnote{
The classical consideration in \cite{bgl} implies that (\ref{ans}) is indeed
the most general quartic ansatz compatible with the $\cN{=}4$ superconformal
algebra.}
\be\label{ans}
V \= V_B(x)\ -\
U_{IJ}(x) \langle \p^I_\a {\bar\p}^{J\a} \rangle\ +\
\sfrac14 F_{IJKL}(x) \langle\p^I_\a\p^{J\a}\bar\p^{K\b}\bar\p^L_\b\rangle\ ,
\ee
with completely symmetric unknown functions $U_{IJ}$ and $F_{IJKL}$
homogeneous of degree $-2$ in $x\equiv\{x^1,\ldots,x^{n+1}\}$.
Here, the symbol $\langle\dots\rangle$ stands for symmetric (or Weyl) ordering.
The ordering ambiguity present in the fermionic sector affects 
the bosonic potential~$V_B$. In contrast to the $\cN{=}2$ superconformal 
extensions~\cite{fm,glp1}, the quartic term is needed, and so we get
\be\label{Qform}
\begin{aligned}
Q_\a &\= \bigl(p_J-\ic\,x^I\,U_{IJ}(x)\bigr)\,\p_\a^J \ \,- 
\sfrac{\ic}{2}\,x^I\,F_{IJKL}(x)\,\<\p^J_\b\,\p^{K\b}\bar\p^L_\a\> \ ,\\[6pt]
\bar Q^\a &\= \bigl(p_J+\ic\,x^I\,U_{IJ}(x)\bigr)\,\bar\p^{J\a} \,-
\sfrac{\ic}{2}\,x^I\,F_{IJKL}(x)\,\<\p^{J\a}\bar\p^{K\b}\bar\p^L_\b\>\ .
\end{aligned}
\ee
To summarize, in order to close the algebra~(\ref{algebra}),
the $D$, $K$, $J_a$, $S_\a$ and $\bar S^\a$ generators remain free,
while $Q_\a$ and $\bar Q^\a$ as well as $H$ acquire corrections as above.
\vspace{1cm}

\section{The structure equations}
\noindent
Inserting the form (\ref{QSfree})--(\ref{ans}) into the 
algebra~(\ref{algebra}), one produces a fairly long list of constraints 
on the potential~$V$. One of the consequences is that~\cite{wyl,bgl,glp2}
\be\label{pot}
U_{IJ} \= \pa_I\pa_J U \und F_{IJKL} \= \pa_I\pa_J\pa_K\pa_L F\ ,
\ee
which introduces two scalar prepotentials. 
The constraints then turn into the following system of 
nonlinear partial differential equations~\cite{bgl,glp2},
\bea
(\pa_I\pa_K\pa_P F)(\pa_J\pa_L\pa_P F)\=
(\pa_J\pa_K\pa_P F)(\pa_I\pa_L\pa_P F)\quad,\qquad
x^I \partial_I \partial_J \partial_K F\=-\de_{JK}\ ,
\label{w1}
\\[6pt]
\pa_I\pa_J U -(\pa_I\pa_J\pa_K F)\,\pa_K U\=0\quad,\qquad
\qquad\qquad\qquad\qquad x^I \pa_I U\=-C\ , \qquad\quad{}
\label{w2}
\eea
which we refer to as the `structure equations'.\footnote{
Wyllard~\cite{wyl} obtained equivalent equations, but employed a different
fermionic ordering.}
Notice that these equations are quadratic in~$F$ but only linear in~$U$.
They are invariant under SO($n{+}1$) coordinate transformations.
The first of~(\ref{w1}) is a kind of zero-curvature condition for a
connection~$\pa^3F$. It coincides with the (generalized) WDVV equation
known from topological field theory~\cite{w,dvv}. The first of~(\ref{w2})
is a kind of covariant constancy for~$\pa U$ in the $\pa^3F$ background.
Since its integrability implies the WDVV~equation projected onto~$\pa U$,
we call it the `flatness condition'. 

The right equations in (\ref{w1}) and~(\ref{w2}) represent homogeneity 
conditions for $U$ and~$F$. They are inhomogeneous with constants 
$\de_{jk}$ and $C$ (the central charge) on the right-hand side
and display an explicit coordinate dependence.
Furthermore, the second equation in (\ref{w1}) can be integrated twice
to obtain
\be\label{w3}
x^I \pa_I F -2F +\sfrac{1}{2} x^I x^I \=0\ ,
\ee
where we used the freedom in the definition of $F$ to put the integration
constants -- a linear function on the right-hand side -- to zero.
It is important to realize that the inhomogeneous term in this integrated
equation excludes the trivial solution $F=0$ equivalent to a homogeneous 
quadratic polynomial. This effect is absent in $\cN{=}2$ superconformal models,
where the four-fermion potential term is not required and, hence, 
$F$ need not appear~\cite{glp1}. This issue is also discussed in~\cite{wyl}.

To simplify the analysis of the structure equations, it is convenient to
separate the center-of-mass and the relative motion of the particles.
This is achieved by a rotation of the coordinate frame,
\be
\{\,x^I\,\} \ \longrightarrow\ \{\,x^i\,,\,X\,\}
\qquad\textrm{with}\qquad i=1,\ldots,n \und
X \= \sfrac{1}{\sqrt{n+1}}\,{\textstyle\sum}_{I=1}^{n+1}\, x^I\ ,
\ee
which introduces relative-motion coordinates $x^i$ for the hyperplane
orthogonal to the center-of-mass direction.
The structure equations then hold for both sets of coordinates independently,
with an accompanying split of the prepotentials and the central charge,
\be
F \= F_{\text{com}}(X)+F_{\text{rel}}(x)\ ,\qquad
U \= U_{\text{com}}(X)+U_{\text{rel}}(x) \und
C \= C_{\text{com}}+C_{\text{rel}}\ ,
\ee
where now $x\equiv\{x^i\}$.
For the center-of-mass coordinate, the solution is trivial:
\be
F_{\text{com}} \= -\sfrac12\,X^2\,\ln|X| \und 
U_{\text{com}} \= -C_{\text{com}}\,\ln|X| \ .
\ee
For the relative coordinates, we simply replace $I,J,\ldots$ by $i,j,\ldots$
and $C\to C_{\text{rel}}$ in the structure equations. 
In the following, we shall investigate the construction of~$F_{\text{rel}}$
and $U_{\text{rel}}$ only and therefore drop the label `rel' from now on.
However, since these coordinates often obscure a permutation invariance
for identical particles, it can be useful to go back to the original~$x^I$
by embedding $\R^n$ into~$\R^{n+1}$ as the hyperplane orthogonal to the
vector $\rho=\sfrac{1}{\sqrt{n+1}}(1,1,\ldots,1)$ for achieving a manifestly 
permutation-symmetric description of the $(n{+}1)$-particle system.
Furthermore, the center-of-mass case is still covered in our analysis by
just taking $n{=}1$.

There are some dependencies among the equations (\ref{w1}) and~(\ref{w2}),
now reduced to the relative coordinates.
The contraction of two left equations with $x^i$ is a consequence of the
two right equations, and therefore only the components orthogonal to~$x$
are independent, effectively reducing the dimension to~$n{-}1$.
This means that only $\sfrac{1}{12}n(n{-}1)^2(n{-}2)$ WDVV equations 
need to be solved and only $\sfrac12n(n{-}1)$ flatness
conditions have to be checked. For $n{=}2$ in particular, the single
WDVV~equation follows from the homogeneity condition in~(\ref{w1}), and
the three flatness conditions are all equivalent. Hence, the nonlinearity
of the structure equations becomes relevant only for $n{\ge}3$.

The scalars $U$ and~$F$ govern the $\cN{=}4$ superconformal extension.
Note, however, that $F$ is defined modulo a quadratic polynomial while $U$
is defined up to a constant.
Together, they determine $V_B$ as\footnote{
We have restored $\hbar$ in the potential to illustrate that 
the $F$ contribution disappears classically.}
\be\label{bc}
V_B \= \sfrac12\,(\pa_iU)(\pa_iU) \ +\ 
\sfrac{\hbar^2}8\,(\pa_i\pa_j\pa_kF)(\pa_i\pa_j\pa_kF) \ .
\ee
We note that $U{\equiv}0$ still yields nontrivial quantum models,
whose potential only vanishes classically.
Finally, from the two right equations in (\ref{w1}) and~(\ref{w2})
it follows that
\be 
x^i\,F_{ijkl} \= -\pa_j\pa_k\pa_lF \und x^i\,U_{ij} \= -\pa_jU\ ,
\ee
which is relevant for~(\ref{Qform}).
\vspace{1cm}

\section{Prepotential ansatz and consequences}
\noindent
Our attack on (\ref{w1}) and~(\ref{w2}) begins with the homogeneity conditions
\be\label{w4}
(x^i\pa_i - 2) F \= -\sfrac12\,x^ix^i \und x^i\pa_i U \= -C\ .
\ee
The general solution to~(\ref{w4}) may be written as
\be\label{Qansatz}
F \= -\sfrac14\sum_s f_s\ Q_s(x)\,\ln|Q_s(x)| \ +\ F_{\rm hom}\und
U \= -\sfrac12\sum_s g_s\,\ln|Q_s(x)|\ +\ U_{\rm hom}
\ee
with quadratic forms $Q_s(x)$, real coefficients $f_s$ and~$g_s$, as well as
homogeneous functions $F_{\rm hom}$ and $U_{\rm hom}$ of degree two and zero,
respectively. The conditions~(\ref{w4}) are obeyed if
\be
Q_s(x) = \sum_{i,j}q^s_{ij}\,x^ix^j \qquad\textrm{satisfies}\qquad
\sum_s f_s\,Q_s(x) = x^ix^i \quad\textrm{and}\quad \sum_s g_s = C\ .
\ee
Unfortunately, it is hard to analyze the WDVV equation~(\ref{w1})
and the flatness condition~(\ref{w2}) in this generality.
Therefore, we take the simplifying ansatz that the quadratic forms
are either of rank one or proportional to the identity form,\footnote{
Our configuration space $\R^n$ carries the Euclidean metric $(\de_{ij})$, 
hence index position is immaterial.}
\be
Q_\a(x) \= \a_i\a_j\,x^ix^j \ =:\ (\ax)^2 \und
Q_R(x)  \= x^ix^i \ =:\ R^2\ ,
\ee
which defines a set $\{\a\}$ of $p$ covectors
\be
\a\=(\a_1,\a_2,\ldots,\a_n) \qquad\textrm{with values}\qquad 
\a(x)\=\ax\=\a_ix^i\ .
\ee
Replacing the label `$s$' by the covector name `$\a$' or by `$R$',
the prepotentials~(\ref{Qansatz}) read
\be\label{Fansatz}
\begin{aligned}
F &\= -\sfrac12\sum_{\a} f_\a\ (\ax)^2\,\ln|\ax|
\ -\ \sfrac12 f_R\,R^2\,\ln R \ +\ F_{\rm hom}(x)\ ,\\
U &\= -\sum_{\a} g_\a\,\ln|\ax| \ -\ g_R\,\ln R \ +\ U_{\rm hom}(x)\ .
\end{aligned}
\ee
The covector part of this ansatz is well known~\cite{wyl,glp2,magra,vese},
but the `radial' terms (labelled `$R$') are new and will be important
for admitting nontrivial solutions~$U$.

The expressions above are invariant under individual sign flips $\a\to-\a$
for each covector, and so we exclude $-\a$ from the set. 
For identical particles our relative configuration space carries an
$n$-dimensional representation of the permutation group~${\cal S}_{n+1}$,
whose action must leave the set $\{\pm\a\}$ invariant.
Furthermore, the $f_\a$ and $g_\a$ couplings have to be constant 
along each ${\cal S}_{n+1}$~orbit.
Finally, a rescaling of~$\ax$ may be absorbed into a renormalization of~$f_\a$.
Therefore, only the rays $\R_+\a$ are invariant data.
We cannot, however, change the sign of~$f_\a$ in this manner.

Compatibility of (\ref{Fansatz}) with the conditions~(\ref{w4}) directly yields
\be\label{hcond}
\sum_\a f_\a\,\a_i \a_j \ +\ f_R\,\de_{ij} \= \de_{ij} \und
\sum_\a g_\a \ +\ g_R \= C\ .
\ee
The second relation fixes the central charge, and the $g_\a$ are independent
free couplings if not forced to zero.
The first relation amounts to a decomposition of $(1{-}f_R)\de_{ij}$
into (usually non-orthogonal) rank-one projectors and imposes 
$\sfrac12n(n{+}1)$ relations on the coefficients~$\{f_\a,f_R\}$ for a given 
set~$\{\a\}$.

All known solutions to the WDVV equations can be cast into the form
(\ref{Fansatz}) with $F_{\rm hom}\equiv0$, so from now on we drop this term.
{}From (\ref{Fansatz}) we then derive
\be\label{YWform}
\begin{aligned}
\pa_i\pa_j\pa_kF &\= -\sum_{\a} f_\a\,\frac{\a_i \a_j \a_k}{\ax} 
\ -\ f_R\,\Bigl\{ \frac{x_i\de_{jk}+x_j\de_{ki}+x_k\de_{ij}}{R^2}
- 2 \frac{x_i x_j x_k}{R^4} \Bigr\} \ , \\[6pt]
\pa_iU &\= -\sum_{\a} g_\a\,\frac{\a_i}{\ax}\ -\ g_R\,\frac{x_i}{R^2} 
\ +\ \pa_iU_{\rm hom}\ ,
\end{aligned}
\ee
and so the bosonic part of the potential takes the form
\be \label{VB}
\begin{aligned}
V_B \=\ &\sfrac12 \sum_{\a,\b} \frac{\a{\cdot}\b}{\ax\;\bx}\,
\Bigl( g_\a g_\b + \sfrac{\hbar^2}4 f_\a f_\b\,(\a{\cdot}\b)^2 \Bigr)
\ -\ \sum_\a g_\a\,\frac{\a_i}{\ax}\,\pa_iU_{\rm hom} \\[4pt]
& +\ \sfrac12\,\frac{1}{R^2}
\Bigl( g_R(2C{-}g_R)+\sfrac{\hbar^2}4(3n{-}2)f_R(2{-}f_R) \Bigr)\ +\
\sfrac12 (\pa_iU_{\rm hom})(\pa_iU_{\rm hom})\ .
\end{aligned}
\ee

The WDVV equation in (\ref{w1}) becomes
\be \label{FF}
\sfrac12\sum_{\a,\b}f_\a f_\b\,\frac{\a{\cdot}\b}{\ax\,\bx}\,
(\a\wedge\b)^{\otimes2}\ +\ f_R\,(2{-}f_R)\,\frac{T}{R^2} \=0
\ee
\be
\begin{aligned}
&\textrm{with}\qquad
(\a\wedge\b)^{\otimes2}_{ijkl} \= (\a_i\b_j-\a_j\b_i)(\a_k\b_l-\a_l\b_k)
\und 
\\[6pt]
&T_{ijkl} \= \de_{ik}\de_{jl}-\de_{il}\de_{jk}-\de_{ik}\hx_j\hx_l
+\de_{il}\hx_j\hx_k-\de_{jl}\hx_i\hx_k+\de_{jk}\hx_i\hx_l
\quad\textrm{where}\quad \hx_i\equiv\sfrac{x_i}{R}\ .
\end{aligned}
\ee
The different singular loci of the various terms in (\ref{FF}) allow one
to separate them, thus
\be \label{FF2}
\smash{\sum_{\a,\b\atop(\a\neq\b)}}\!f_\a f_\b\,
\frac{\a{\cdot}\b}{\ax\,\bx}\,(\a\wedge\b)^{\otimes2} \= 0 \und
f_R\,(2{-}f_R) \= 0\ .
\ee
The two admissible choices for $f_R$, 
\be \label{flip}
f_R= 0 \quad{\buildrel(\ref{hcond})\over\longrightarrow}\quad
\sum_\a f_\a \a{\otimes}\a =  \unity
\qquad\textrm{or}\qquad 
f_R= 2 \quad{\buildrel(\ref{hcond})\over\longrightarrow}\quad 
\sum_\a f_\a \a{\otimes}\a = -\unity\ ,
\ee
are related by flipping the signs of all coefficients~$f_\a$, 
i.e.~$f_\a\to-f_\a$. Note, however, that the $n{=}2$ case is special,
since then $T{\equiv}0$ and (\ref{FF}) is identically satisfied,
so no restrictions on $f_R$ arise.

The flatness condition in (\ref{w2}), on the other hand, 
is already nontrivial at $n{=}2$ and reads
\be \label{UF} 
\pa_i\pa_j U\ +\ \sum_\a f_\a\frac{\a_i\a_j}{\ax}\,\a{\cdot}\pa U\ +\ f_R\,
\Bigl\{\frac{x_i\pa_jU+x_j\pa_iU-\de_{ij}C}{R^2}+\frac{2x_ix_jC}{R^4}\Bigr\}
\=0 \ .
\ee
In particular, its trace,
\be \label{Utrace}
\pa{\cdot}\pa\,U \ +\ \sum_\a f_\a \frac{\a{\cdot}\a}{\ax} \a{\cdot}\pa U \=
C\,f_R\,\frac{n}{R^2}\ ,
\ee
and its projection onto some covector~$\b$,
\be \label{Uproj}
(\b{\cdot}\pa)^2 U\ +\ \sum_\a f_\a \frac{(\a{\cdot}\b)^2}{\ax} \a{\cdot}\pa U
\ +\ 2f_R\frac{\bx}{R^2}\,\b{\cdot}\pa U \= 
C\,f_R \Bigl( \frac{\b{\cdot}\b}{R^2}-\frac{2(\bx)^2}{R^4}\Bigr)\ ,
\ee
prove to be useful.
They are potentially singular at $R{=}0$ and on the hyperplanes $\ax{=}0$.
For example, near $\bx{=}0$ (but away from $R{=}0$) we may approximate
(\ref{Uproj}) by 
\be
(\b{\cdot}\pa)^2U\ +\ \frac{f_\b\,\b{\cdot}\b}{\bx}\,\b{\cdot}\pa U\ \approx\ 0
\qquad{\buildrel f_\b\ge-1\over\longrightarrow}\qquad
U\ \sim\ (\bx)^{1-f_\b} \quad\textrm{for}\quad \bx\sim0\ ,
\ee
which displays the leading singularity structure of~$U$ (and thus of $V_B$)
on the $\bx{=}0$ hyperplane provided that $f_\b$ is sufficiently large.

Of course, there is always the trivial $C{=}0$ solution, which puts
$g_R=g_\a=0\ \ \forall\a$.
As long as we keep $U_{\rm hom}$ to be nonzero, it is not too illuminating
to insert the covector expression~(\ref{YWform}) into the above equations. 
So let us, for a moment, 
ponder the consequences of putting $U_{\rm hom}\equiv0$ in~(\ref{Fansatz}). 
In such a case for $n{>}2$, 
(\ref{UF}) together with (\ref{YWform}) implies 
\be \label{UF2}
g_\a\,(1{-}\a{\cdot}\a f_\a)\=0 \ ,\quad
\smash{\sum_{\a,\b\atop(\a\neq\b)}}\!
g_\a f_\b\,\frac{\a{\cdot}\b\ \b_i\b_j}{\ax\ \bx} \= 0 \ ,\quad
g_\a\,f_R \= C\,f_R \= g_R \= 0 \ ,
\ee
which essentially kills all radial terms and fixes 
$f_\a=\sfrac{1}{\a{\cdot}\a}$ unless $g_\a=0$.
Turning on all $g_\a$ would then saturate the first option in~(\ref{flip}),
\be \label{sumrule}
\sum_\a\frac{\a{\otimes}\a}{\a{\cdot}\a}\=\unity 
\qquad\longrightarrow\qquad \a{\cdot}\b=0\ \ \forall\a,\b\ ,
\ee
because this partition of unity is an orthonormal one and the number~$p$
of covectors~$\a$ must be equal to~$n$.
Clearly, such a system is reducible:
If a set of covectors decomposes into mutually orthogonal subsets,
(\ref{FF}) and~(\ref{UF}) -- at $f_R{=}0{=}g_R$ -- 
hold for each subset individually.
Then, the partial prepotentials just add up to the total $F$ or~$U$.
In fact, we have already encountered such a decomposition when separating
the center-of-mass degree of freedom.
Here, however, it is the {\it relative\/} motion of the particles which 
can be factored into independent parts. Since the irreducible
relative-particle systems are the building blocks for all models,
the case of $p=n$ is just a collection of $n{=}1$ systems and does
not provide an interesting solution.
We learn that $U_{\rm hom}{\equiv}0$ is not an option for $n{>}2$.

Let us finally take a look at the special case of $n{=}2$, 
i.e~relative motion in a three-particle system.
First, as already mentioned, the $n{=}2$ WDVV~equation is empty; 
it follows from~(\ref{hcond}),
which can be fulfilled for any set of more than one covector.
Hence, $f_R$ is unrestricted.
Second, at $n{=}2$ the content 
of~(\ref{UF}) is fully captured by its trace~(\ref{Utrace}), which 
in this case allows nontrivial solutions even with $U_{\rm hom}\equiv0$.
Namely, inserting the second line of~(\ref{YWform}) with $U_{\rm hom}\equiv0$
into~(\ref{Utrace}) one obtains
\be\label{trace} 
\sum_\a g_\a\,(1{-}\a{\cdot}\a f_\a)\,\frac{\a{\cdot}\a}{(\ax)^2}\ -\
\smash{\sum_{\a,\b\atop(\a\neq\b)}}\!
g_\a f_\b\,\frac{\a{\cdot}\b\ \b{\cdot}\b}{\ax\ \bx}\ -\
\frac{1}{R^2} \Bigl( 2(n{-}1)g_R + n(C{-}g_R)f_R \Bigr) \= 0 \ ,
\ee
which splits into
\be \label{UF3}
g_\a\,(1{-}\a{\cdot}\a f_\a)\=0 \ ,\qquad
\smash{\sum_{\a,\b\atop(\a\neq\b)}}\!
g_\a f_\b\,\frac{\a{\cdot}\b\ \b{\cdot}\b}{\ax\ \bx}\=0 \ ,\qquad
g_R \=\sfrac{f_R}{f_R-2+2/n}\,C \ .
\ee
If all couplings~$g_\a$ are nonzero, then
\bea \label{contra}
f_\a\=\sfrac{1}{\a{\cdot}\a}\ >0
\qquad{\buildrel(\ref{hcond})\over\longrightarrow}\qquad
f_R \= 1-\sfrac{p}{n} \und
g_R \= \sfrac{p-n}{p+n-2}\,C \quad{} \\[6pt] 
\label{besides} \textrm{besides} \qquad
\sum_\a \frac{\a{\otimes}\a}{\a{\cdot}\a} \= \sfrac{p}{n}\,\unity \und
\smash{\sum_{\a,\b\atop(\a\neq\b)}}\!g_\a\;\frac{\a{\cdot}\b}{\ax\ \bx} \= 0\ .
\eea
These equations will be analyzed in Section~8.
We already see that the radial terms are essential for having $p>n$.
Of course, we are to put $n{=}2$ in the equations above, but
we have displayed the general formulae to make explicit the conflict
between (\ref{contra}) and~(\ref{FF2}) for $n{\ge}3$ and $p{>}n$,
which essentially rules out $U_{\rm hom}{\equiv}0$ solutions beyond $n{=}2$.
\vspace{1cm}

\section{$U{=}0$ solutions: root systems}
\noindent
The obvious strategy for solving the structure equations 
is to first construct a prepotential~$F$ satisfying~(\ref{w1}),
i.e.~find covectors~$\a$ (and coefficients $f_\a$) subject to
(\ref{flip}) and~(\ref{FF2}). Without loss of generality we restrict ourselves 
to the first of the two cases in~(\ref{flip}) and put~$f_R{=}0$.
The structure equations are linear in the prepotential~$U$, 
and so a solution to the WDVV~equation trivially extends to a full solution 
$(F,U{\equiv}0)$ for $C{=}0$.

In 1999, Martini and Gragert~\cite{magra} discovered that, 
in~(\ref{Fansatz}) with $f_R{=}0{=}g_R$, 
taking $\{\a\}$ to be a (positive) root system of any simple 
Lie algebra yields a valid prepotential~$F$. 
Shortly thereafter, it was proved~\cite{vese} that certain deformations
of root systems are also allowed, as well as the root systems of {\it any\/}
finite reflection group, thus adding the non-crystallographic Coxeter groups
to the list. In the following, we shall rederive these results and generalize
them.

Let us begin with the simply-laced root systems. Here, any two positive
roots $\a$ and~$\b$ are either orthogonal, or else add or subtract to
another positive root, then giving rise to an equilateral triangle
\be
\a+\b+\g \= 0 \qquad\longrightarrow\qquad 
\a\wedge\b \= \b\wedge\g \= \g\wedge\a \und
\a{\cdot}\b \= \b{\cdot}\g \= \g{\cdot}\a\ .
\ee
The contribution of the pairs $(\a,\b)$, $(\b,\g)$ and $(\g,\a)$ 
to~(\ref{FF2}) is thus proportional to
\be \label{id1}
\frac{f_\a\,f_\b}{\ax\;\bx}\ +\
\frac{f_\b\,f_\g}{\bx\;\gx}\ +\
\frac{f_\g\,f_\a}{\gx\;\ax}\ ,
\ee
which vanishes precisely when $f_\a=f_\b=f_\g$.
We recognize the triple $(\a,\b,-\g)$ as the positive roots of~$A_2$.

It is not hard to see that in~(\ref{FF2}) the sum over all non-orthogonal pairs 
$(\a,\b)$ of positive $ADE$ roots can be decomposed into partial sums over the 
three pairs of a triple. Two triples may share a single root but not a pair.
Since all triples are connected in this way, all $f_\a$ are equal,\footnote{
The trivial way to avoid this conclusion puts $f_\a{=}0$ for sufficiently
many roots such that the system decomposes into mutually orthogonal parts,
with their $f_\a$ values determined individually via~(\ref{flip}).}
and their value is fixed by the homogeneity condition~(\ref{flip}),
which implies that our root system must be of rank~$n$.
To find~$f$, recall that, for any Lie algebra and with $\a{\cdot}\a{=}2$
for the long roots, one has
\be
\sum_{\a\in\Phi^+} \a\otimes\a \= h^\vee\,\unity \und
\sum_{\a\in\Phi^+} 2\,\frac{\a\otimes\a}{\a\cdot\a} \= h\,\unity\ ,
\ee
where $\Phi^+$ is the set of positive roots, and $h$ and $h^\vee$ denote 
the Coxeter and dual Coxeter numbers, respectively. 
Thus, $f=1/h^\vee$ in the $ADE$ case, where $h{=}h^\vee$.
\begin{center}
\begin{tabular}{|l|ccccccccc|ccc|}
\hline
$\Phi^+$ & $A_n$ & $B_n$ & $C_n$ & $D_n$ & $E_6$ & $E_7$ & $E_8$ & $F_4$ &$G_2$
         & $H_3$ & $H_4$ & $I_2(p)$ \\
\hline & & & & & & & & & & & & \\[-10pt]
$h$      & $n{+}1$ & $2n$ & $2n$ & $2n{-}2$ & 12 & 18 & 30 & 12 & 6 
         & 10 & 30 & $p$ \\[4pt]
$h^\vee$ & $n{+}1$ & $2n{-}1$ & $n{+}1$ & $2n{-}2$ & 12 & 18 & 30 & 9 & 4 
         & -- & -- & --  \\
\hline
\end{tabular}
\end{center}

In essence, the root systems of all $ADE$~Lie algebras provide us with
prepotentials~\cite{magra}
\be \label{fade}
F_{ADE} \= -\sfrac{1}{2h^\vee} \sum_{\a\in\Phi^+} (\ax)^2\,\ln|\ax|\ .
\ee

What about the other root systems?
There, we have long roots, with length${}^2=2$, and short roots,
with length${}^2=2/r$, where $r=2$ or $3$. 
Any two non-orthogonal short roots add or subtract to another short root,
and the same is true for the long roots. Hence, for the short/short or 
long/long pairs in our double sum we can again employ (\ref{id1})=0,
which identifies the $f$~coefficients in each triple.
However, we also encounter long/short pairs in~(\ref{FF2}).
The key is to realize that the $ADE$ triple $(\a,\b,\a{+}\b)$ represents the
$\b$-string of roots through~$\a$. The root string concept works for any pair
of roots and in general groups together $r{+}2$ coplanar roots
$(\a,\b,\a{+}\b,\a{+}2\b,\ldots,\a{+}r\b)$, 
with $\a$ being long, $\b$ short and $\a\cdot\b=-1$.
\begin{figure}[ht]
\centerline{\includegraphics[width=16cm]{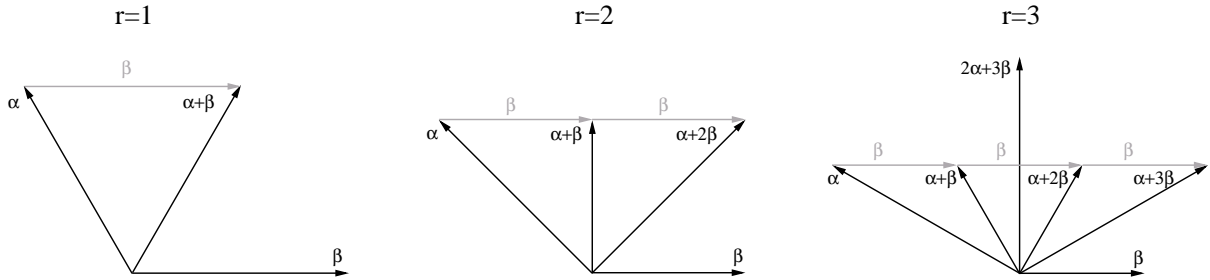}}
\caption{Short-root strings through a long root, 
for length${}^2$ ratios $r=1,2,3$}
\label{fig:5}
\end{figure}

For the long/short pairs in $B_n$, $C_n$ and $F_4$ ($r{=}2$) 
the role of (\ref{id1})=0 is then taken by a four-root identity 
based on the quadruple $(\a,\b,\a{+}\b,\a{+}2\b)$.
With scalar products
\begin{center}
\begin{tabular}{|c|cc|cc|}
\hline
$\cdot$    & $\ \b$ & $\a{+}\b$ & $\ \a$ & $\a{+}2\b$ \\
\hline & & & & \\[-10pt]
$\b$       & \  1 & 0 & \ -1 & 1 \\[4pt]
$\a{+}\b$  & \  0 & 1 & \  1 & 1 \\[4pt]
\hline & & & & \\[-10pt]
$\a$       & \ -1 & 1 & \  2 & 0 \\[4pt]
$\a{+}2\b$ & \  1 & 1 & \  0 & 2 \\
\hline
\end{tabular}
\end{center}
and the relevant wedge products all equal modulo sign, 
the equal-length pairs drop out, and
the quadruple yields just four long/short pairs 
for the sum in~(\ref{FF2}),
\be \label{id2}
-\ \frac{f_\a\,f_\b}{\ax\;\bx}\
+\ \frac{f_\a\,f_{\a+\b}}{\ax\;(\a{+}\b){\cdot}x}\
+\ \frac{f_{\a+2\b}\,f_\b}{(\a{+}2\b){\cdot}x\;\bx}\ 
+\ \frac{f_{\a+2\b}\,f_{\a+\b}}{(\a{+}2\b){\cdot}x\;(\a{+}\b){\cdot}x}\ .
\ee
This expression vanishes only when it must, namely for
\be
f_\a \= f_{\a+2\b} \ =:\ f_{\textrm{L}} \und
f_\b \= f_{\a+\b}  \ =:\ f_{\textrm{S}} \ .
\ee

Like in the $ADE$ case, each non-orthogonal pair of roots defines
a unique plane, which carries either a triple or a quadruple. Hence,
the sum in~(\ref{FF2}) again splits into sums over the pairs of a triple
or a quadruple, which yield zero individually. Since each plane shares
its roots with other planes and all are connected unless the system is
decomposable, all long roots come with the same coefficient
$f_{\textrm{L}}$, and all short roots with~$f_{\textrm{S}}$.
The normalization in~(\ref{flip}) then reads
\be \label{fnorm}
f_{\textrm{L}} \sum_{\a\in\Phi^+_{\textrm{L}}} \a\otimes\a \ +\
f_{\textrm{S}} \sum_{\a\in\Phi^+_{\textrm{S}}} \a\otimes\a \=\unity\ ,
\ee
where $\Phi^+_{\textrm{L}}$ and $\Phi^+_{\textrm{S}}$ stand for the
positive long and short roots, respectively. Using
\be
\sum_{\a\in\Phi^+_{\textrm{L}}} \a\otimes\a 
\= \frac{rh^\vee{-}h}{r{-}1}\,\unity \und
\sum_{\a\in\Phi^+_{\textrm{S}}} \a\otimes\a 
\= \frac{h{-}h^\vee}{r{-}1}\,\unity \ ,
\ee
the solution to~(\ref{fnorm}) is a one-parameter family,
\be \label{ffamily}
\begin{aligned}
f_{\textrm{L}} &\= \sfrac{1}{h^\vee} + (h{-}h^\vee)t 
\ \= \sfrac{1}{h} + (h{-}h^\vee)t' \ , \\
f_{\textrm{S}} &\= \sfrac{1}{h^\vee} + (h{-}rh^\vee)t 
\= r\,\bigl\{\sfrac{1}{h} + (\sfrac{h}{r}{-}h^\vee)t'\bigr\}\ ,
\end{aligned}
\ee
with $t=t'-\frac{1}{hh^\vee}\in\R$.
Therefore, we arrive at a {\it family\/} of prepotentials
\be \label{Fsol}
F \= 
-  \sfrac12 f_{\textrm{L}} \sum_{\a\in\Phi^+_{\textrm{L}}}(\ax)^2\,\ln|\ax| \
-\ \sfrac12 f_{\textrm{S}} \sum_{\a\in\Phi^+_{\textrm{S}}}(\ax)^2\,\ln|\ax| \ .
\ee

Incidentally, the formulae (\ref{ffamily}) and~(\ref{Fsol}) 
hold for {\it all\/} root systems, 
including the $ADE$ ($r{=}1$) and $G_2$ ($r{=3}$) cases. 
The only $r{=}3$ example, $G_2$, is trivial since of rank~two,
but let us anyway also prove the assertion for this case.
The six positive roots of~$G_2$ contribute
3 short/short, 3 long/long and 6 long/short pairs to the sum in~(\ref{FF2}).
As argued before, the contributions of the equal-length pairs vanish by virtue
of~(\ref{id1})=0, provided $f_\a{=}f_{\textrm{S}}$ for the short roots and 
$f_\a{=}f_{\textrm{L}}$ for the long ones. The mixed pairs yield
\be \label{id3}
\sfrac{-1}{\ax\,\bx}+\sfrac{1}{\ax\,(\a{+}\b){\cdot}x}+
\sfrac{1}{(\a{+}3\b){\cdot}x\,\bx}+
\sfrac{1}{(\a{+}3\b){\cdot}x\,(\a{+}2\b){\cdot}x}+
\sfrac{1}{(2\a{+}3\b){\cdot}x\,(\a{+}\b){\cdot}x}+
\sfrac{1}{(2\a{+}3\b){\cdot}x\,(\a{+}2\b){\cdot}x}
\ee
for a long root~$\a$ and a short root~$\b$, with $\a{\cdot}\b=-1$,
which as simple roots generate the $G_2$ system. It is quickly verified 
that the above expression indeed vanishes, which proves our claim.
Hence, for all Lie-algebra root systems, we have proved the identity
\be \label{rootid}
\smash{\sum_{\a,\b\atop(\a\neq\b)}}\!
\frac{\a{\cdot}\b}{\ax\ \bx} \= 0  \qquad\textrm{for}\qquad
(\a,\b)\ \in\ \bigl(\Phi^+_{\textrm{L}},\Phi^+_{\textrm{L}}\bigr)
\ \textrm{or}\ \bigl(\Phi^+_{\textrm{S}},\Phi^+_{\textrm{S}}\bigr)
\ \textrm{or}\ \bigl(\Phi^+_{\textrm{L}},\Phi^+_{\textrm{S}}\bigr)\ ,
\ee
which is effectively equivalent to the WDVV equation.
Our solution~(\ref{ffamily}) for the $f$~coefficients generalizes the one 
of~\cite{magra,vese} and reduces to them at $t{=}0$. 
One might think that the one-parameter freedom is ficticious 
since $f_{\textrm{L}}$ and $f_{\textrm{S}}$ may be absorbed into 
the roots. However, this is not so because $f_{\textrm{L}}$ and 
$f_{\textrm{S}}$ may have opposite signs, which is crucial for constructing
$U$~solutions in this $F$~background.

We have also checked the non-crystallographic
Coxeter groups $H_3$, $H_4$ and $I_2(p)$ for $p{=}5$ and $p{>}6$.\footnote{
Up to a root rescaling, \ $I_2(2)=A_1{\oplus}A_1$, \ $I_2(3)=A_2$, \
$I_2(4)=B_2$ or~$C_2$, \ and $I_2(6)=G_2$.}
Of these, the dihedral $I_2$ series
\be \label{Iroots}
\{\a\} \= \{ \ \cos(k\pi/p)\,e_1 +\,\sin(k\pi/p)\,e_2\
|\ k=0,1,\ldots,p{-}1\ \}
\ee
trivially fulfils~(\ref{FF}), as it is of rank~two.
\begin{figure}[ht]
\centerline{\includegraphics[width=16cm]{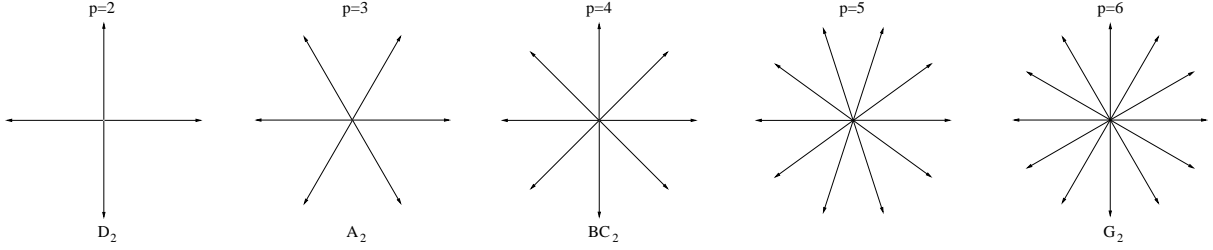}}
\caption{Root systems of the dihedral groups $I_2(p)$ for $p=2,3,4,5,6$}
\label{fig:6}
\end{figure}
\vspace{1cm}

\section{$U{=}0$ solutions: orthocentric simplices}
\noindent
In order to generalize the root-system solutions found in the
previous section, in this section we take a more general look at 
the $n{=}3$ case. Again, the goal is to solve the WDVV equation~(\ref{FF2})
and the homogeneity condition~(\ref{flip}) for $f_R{=}0$.

Previously we have mentioned that any set of $p{\ge}2$ covectors
in $n{=}2$ dimensions solves~(\ref{w1}), because the WDVV equation is empty
and (\ref{hcond}) only serves to restrict $f_\a$ and~$f_R$.
We now deliver a simple argument.
Let us represent a covector $\a\in\R^2$ by a complex number~$a\in\C$. Then,
the traceless and the trace part of the homogeneity condition~(\ref{hcond})
translate to
\be \label{hcond2}
\sum_a f_a\,a^2 \= 0 \und \sum_a f_a\,a\bar{a} \= 2\,(1{-}f_R)\ ,
\ee
respectively, where $\bar{a}$ is the complex conjugate of~$a$ and
$f_a\equiv f_\a\in\R$.
Since the length of each covector can be changed by rescaling the
corresponding~$f$, it is evident that for more than one covector
one can always select these coefficients in such a way that the
complex numbers $f_a\,a^2$ form a closed polygonal chain in two dimensions,
thus satisfying the first of~(\ref{hcond2}).
A common rescaling then takes care of the second equation as well,
while $f_R$ can still be dialed at will.
Therefore, by taking the complex square roots of the edge vectors
of any closed polygonal chain, we obtain an admissible set of covectors.

Before moving on to three dimensions, it is instructive to work out
the $f_\a$ coefficients from the homogeneity condition~(\ref{flip})
for $f_R{=}0$ and $n{=}2$.
For the case of two covectors $\{\a,\b\}$, necessarily $\a{\cdot}\b=0$. 
For $p{=}3$ coplanar covectors $\{\a,\b,\g\}$,
the homogeneity condition~(\ref{flip}) uniquely fixes the $f$~coefficients to
\be \label{f2dim}
f_\a \= -\frac{\b\cdot\g}{\a{\wedge}\b\ \g{\wedge}\a}
\qquad\textrm{and cyclic}\ ,
\ee
due to the identity
\be
\b{\wedge}\g\ \b{\cdot}\g\ \a_i\a_j\ +\ \textrm{cyclic} \=
-\a{\wedge}\b\ \b{\wedge}\g\ \g{\wedge}\a\ \de_{ij}\ .
\ee
The traceless part of the homogeneity condition should imply
the single WDVV~equation~(\ref{FF2}) in two dimensions.
Indeed, the choice~(\ref{f2dim}) turns the latter into
\be
\a{\wedge}\b\ \g{\cdot}x\ +\
\b{\wedge}\g\ \a{\cdot}x\ +\
\g{\wedge}\a\ \b{\cdot}x\= 0
\ee
which is identically true.
Without loss of generality we may assume that $\a+\b+\g=0$, 
i.e.~the three covectors form a triangle. 
In this case we have $\a{\wedge}\b=\b{\wedge}\g=\g{\wedge}\a=2A$, 
where the area~$A$ of the triangle may still be scaled to~$\sfrac12$, 
and (\ref{f2dim}) simplifies to
\be \label{f2sim}
f_\a \= -\frac{\b\cdot\g}{4\,A^2}
\qquad\textrm{and cyclic}\ .
\ee
\begin{figure}[ht]
\centerline{\includegraphics[width=8cm]{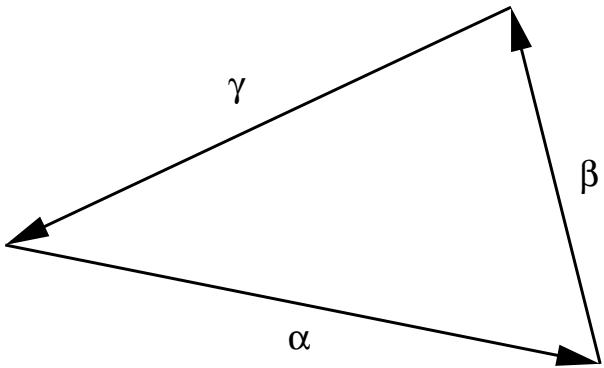}}
\caption{Triangular configuration of covectors}
\label{fig:1}
\end{figure}

In dimension $n{=}3$, the minimal set of three covectors must form
an orthogonal basis, with $f_\a=1/\a{\cdot}\a$. Let us skip the cases
of four and five covectors and go to the situation of $p{=}6$ covectors
because the homogeneity condition~(\ref{flip}) then precisely determines
all $f$~coefficients.
However, it is not true that six generic covectors can be scaled to form
the edges of a polytope. The space of six rays in~$\R^3$ modulo rigid~SO(3)
is nine dimensional, while the space of tetrahedral shapes (modulo size) has
only five dimensions. In order to generalize the $n{=}2$ solution above,
let us assume that our six covectors can be scaled to form a tetrahedron,
with edges $\{\a,\b,\g,\a',\b',\g'\}$ where $\a'$ is dual to~$\a$
and so on.
\begin{figure}[ht]
\centerline{\includegraphics[width=8cm]{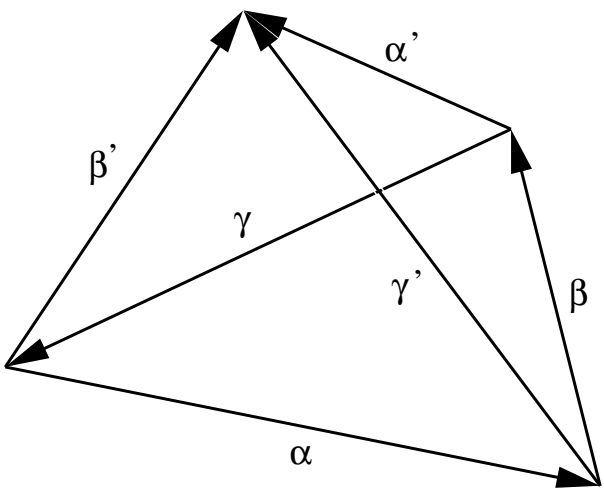}}
\caption{Tetrahedral configuration of covectors}
\label{fig:2}
\end{figure}
Any such tetrahedron is determined by giving three nonplanar
covectors, say $\{\a,\b,\g'\}$, which up to rigid rotation are fixed by
six parameters, corresponding to the shape and size of the tetrahedron.

Let us try employing the triangle result~(\ref{f2sim}) to patch together
the unique solution to the homogeneity condition~(\ref{flip}) for the
tetrahedron. To satisfy the traceless part of the relation, we take the
$f$~coefficients around any face to be proportional to the triangular
ones~(\ref{f2sim}). 
Now each edge is shared by two triangular faces, so we should have
\be
f_\a \= -\la_{\a\b\g}\,\b{\cdot}\g \= -\la_{\a\b'\g'}\,\b'{\cdot}\g'
\ee
and so forth cyclicly around the triangles~$\<\a\b\g\>$ and $\<\a\b'\g'\>$,
with coefficients~$\la_{\cdots}$ depending only on the triangle indicated.
It is then tempting to put
\be \label{tansatz}
f_\a \= -\la\ \b{\cdot}\g\ \b'{\cdot}\g'\ ,\qquad
f_\b \= -\la\ \g{\cdot}\a\ \g'{\cdot}\a'\ ,\qquad
f_\g \= -\la\ \a{\cdot}\b\ \a'{\cdot}\b'
\ee
and so on using the tetrahedral incidences, with $\la$ depending only
on the volume~$V$ of the tetrahedron. However, comparing the two previous
sets of equations we see that this can only work if
\be \label{oc1}
\b'{\cdot}\g' \= \g'{\cdot}\a' \= \a'{\cdot}\b' \= \sfrac{\la_{\a\b\g}}{\la}
\ee
and likewise for any three convergent edges dual to some face.
These eight relations are non-generic but immediately equivalent to
the three conditions
\be \label{oc2}
\a\cdot\a' \=0\ ,\qquad \b\cdot\b' \=0\ ,\qquad \g\cdot\g' \=0
\ee
for the pairs of dual (skew) edges of the tetrahedron.
Such tetrahedra, called `orthocentric'~\cite{ehm}, are characterized by
the fact that all four altitudes are concurrent (in the orthocenter) and
their feet are the orthocenters of the faces. The space of orthocentric
tetrahedra is of codimension two inside the space of all tetrahedra and 
represents a three-parameter deformation of the $A_3$~root system 
(ignoring the overall scale).

For orthocentric tetrahedra, our ansatz~(\ref{tansatz}) is successful:
Due to the identity
\be
\b{\cdot}\g\ \b'{\cdot}\g'\ \a_i\a_j\ +\  
\b{\cdot}\g'\ \b'{\cdot}\g\ {\a'}_i{\a'}_j\ +\
\textrm{cyclic} \= -36\,V^2\,\de_{ij}\ ,
\ee
the homogeneity condition~(\ref{flip}) is obeyed for
\be \label{f3sim}
f_\a \=
-\frac{\b{\cdot}\g\ \b'{\cdot}\g'}{36\,V^2} \und
f_{\a'} \= 
-\frac{\b{\cdot}\g'\ \b'{\cdot}\g}{36\,V^2} 
\ee
plus their cyclic images.
What about the WDVV~equation in this case?
The 15 pairs of edges in the double sum of~(\ref{FF2}) group
into four triples corresponding to the tetrahedron's faces plus
the three skew pairs. Using~(\ref{f3sim}), the contribution of the
$\<\a\b\g\>$~face becomes proportional to
$\b'{\cdot}\g'\,\g{\cdot}x+\textrm{cyclic}$,
which vanishes thanks to~(\ref{oc1}).
Repeating this argument for the other faces, we see that
the concurrent edge pairs do not contribute to the double sum in~(\ref{FF2}),
which leaves us with the three skew pairs. At this point, the orthocentricity
again comes to the rescue via~(\ref{oc2}), and the WDVV~equation is obeyed.
Apparently, any reduction of the WDVV~equation to some face already follows
from the homogeneity condition, and the only independent projection is
associated with the skew edge pairs.

Although we do not know the $f$~coefficients for a general tetrahedron,
we can employ a dimensional reduction argument to prove
that the WDVV equation already enforces the orthocentricity.
Consider the limit $\hat{n}\cdot x\to\infty$ for some 
fixed covector~$\hat{n}$ of unit length. Decomposing
\be \label{limit}
\a \= \a{\cdot}\hat{n}\;\hat{n} + \a_\perp \qquad\longrightarrow\qquad
\ax \= \a{\cdot}\hat{n}\;\hat{n}{\cdot}x + \a_\perp{\cdot}x
\ee
we see that any factor $\frac1\ax$ vanishes in this limit
unless $\a{\cdot}\hat{n}=0$. Thus, only covectors perpendicular to~$\hat{n}$
survive in (\ref{FF}), reducing the system to the hyperplane
orthogonal to~$\hat{n}$. In addition, $\frac1R\to0$ as well, killing all
radial terms in the process.\footnote{
Note, however, that the reduced system in general does not fulfil the
homogeneity conditions~(\ref{hcond}) since the `lost covectors' have
nonzero projections onto the hyperplane.}
In a general tetrahedron, take $\hat{n}$ to point in the direction 
of $\a{\wedge}\a'$.
Then, the limit $\hat{n}{\cdot}x\to\infty$ retains only 
the covectors $\a$ and~$\a'$, and the WDVV equation reduces to a single term,
which vanishes only for $\a{\cdot}\a'=0$. Equivalently, the plane spanned by
$\a$ and $\a'$ contains no further covector, and two covectors in two
dimensions must be orthogonal. The same argument applies to
$\b{\cdot}\b'$ and $\g{\cdot}\g'$, completing the proof.

\begin{figure}[ht]
\centerline{\includegraphics[width=8cm]{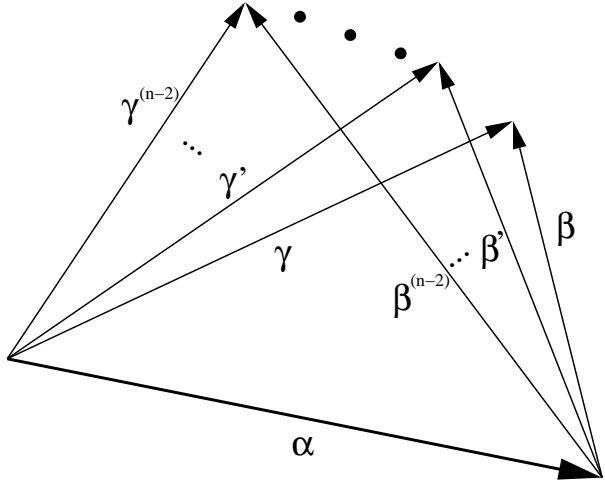}}
\caption{Faces sharing an edge of an $n$-simplex}
\label{fig:3}
\end{figure}
This scheme may be taken to any dimension~$n$. A simplicial configuration
of $\sfrac12n(n{+}1)$ covectors is already determined by $n$ independent
covectors, which modulo~SO($n$) are given by $\sfrac12n(n{+}1)$ parameters.
The homogeneity condition~(\ref{flip}) uniquely fixes the $f$~coefficients.
Employing an iterated dimensional reduction to any plane spanned by a
skew pair of edges and realizing that no other edge lies in such a plane, we
see that the WDVV equation always demands such an edge pair to be orthogonal.
This condition renders the $n$-simplex orthocentric and reduces the
number of degrees of freedom to~$n{+}1$ (now including the overall scale
given by the $n$-volume~$V$). In this situation we can write down
the unique solution to both the homogeneity condition and the WDVV~equation,
\be
f_\a \= \frac{\b{\cdot}\g\ \b'{\cdot}\g'\ \b''{\cdot}\g''\
\cdots\ \b^{(n-2)}{\cdot}\g^{(n-2)}}{(n!\ V)^2}\ ,
\ee
where the edge~$\a$ is shared by the $n{-}1$ faces $\<\a\b\g\>$,
$\<\a\b'\g'\>$, $\ldots$, $\<\a\b^{(n-2)}\g^{(n-2)}\>$, and we have
oriented all edges as pointing away from~$\a$.
This formula works because any sub-simplex, in particular any tetrahedral
building block, is itself orthocentric.
To summarize, the WDVV solutions for simplicial covector configurations
in any dimension are exhausted by an $n$-parameter deformation of the
$A_n$~root system. The $n$~moduli are relative angles and do not include the
$\sfrac12n(n{+}1)$ trivial covector rescalings, which, apart from the common
scale, destroy the tetrahedron. It has to be checked whether our deformation
coincides with the $A_n$~deformation found in~\cite{chaves} in a different
setting.

As a concrete example, the reader is invited to work out the details for
the generic (scaled) orthocentric 4-simplex with vertices
\be
\begin{aligned} &
A: (0,0,0,0) \qquad
B: (1,0,0,0) \qquad
C: (x,y,0,0) \qquad{} \\[8pt] &
D: (x,\sfrac{x(1-x)}{y},z,0) \qquad
E: (x,\sfrac{x(1-x)}{y},\sfrac{x(1-x)(y^2-x(1-x))}{y^2\,z},w) \quad.
\end{aligned}
\ee

\begin{figure}[ht]
\centerline{\includegraphics[width=8cm]{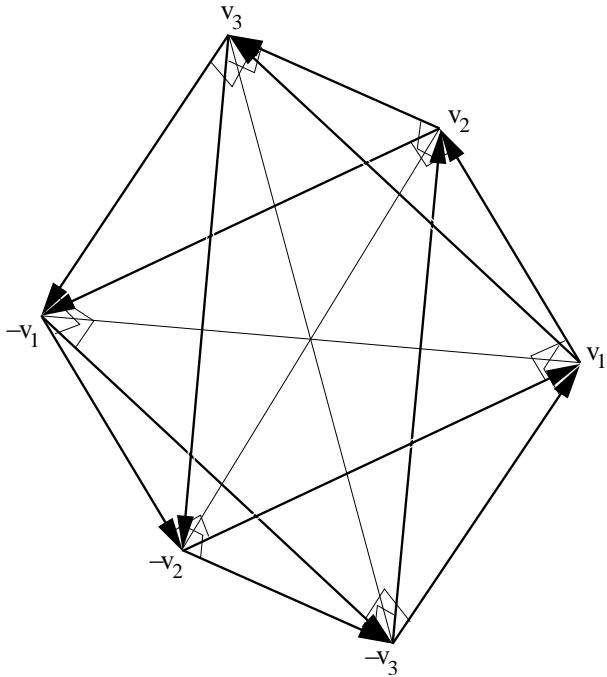}}
\caption{Octahedral configuration of covectors}
\label{fig:4}
\end{figure}
Orthocentric simplices are not the only generalization of our analysis
of six covectors in three dimensions. Recalling that $A_3=D_3$, we know that
the six edges of a regular tetrahedron can be reassembled into one-half of
a regular octahedron. Let us relax the regularity and look at a more general
octahedron defined by six vertices $\pm v_1$, $\pm v_2$ and~$\pm v_3$, 
which are fixed (up to rigid rotations) by six parameters, just like for the
tetrahedron. For the full set of edges we need to include here also the
negatives of all positive covectors,
\be \label{octaroots}
\{\pm\a\} \= \bigl\{ \a_{(\pm i\pm j)} = \pm v_i{\pm}v_j
\quad\textrm{for}\quad 1\le i<j\le 3 \bigr\}\ .
\ee
With $f_{-\a}=f_\a$, 
the homogeneity condition uniquely fixes all $f$~coeficients.
For the WDVV~equation, let us again consider the dimensional reduction to
the plane spanned by any pair of covectors, and restrict to the positive ones.
Like for the tetrahedron, it turns out that such a plane contains either a 
triangular face or just two convergent covectors $\a_{(i+j)}$ and $\a_{(i-j)}$.
The reduced WDVV~equation requires a right angle between the latter,
which puts $v_i{\cdot}v_i=v_j{\cdot}v_j$, and so all three vertices must
have the same distance from the origin.  We are not aware of a particular
name for such octahedra, which admit a circumsphere. In any case, these two 
conditions and ignoring the overall scale reduce the modular space to a 
three-dimensional one, which we already identified as the space of 
orthocentric tetrahedral shapes.

The virtue of this alternative picture is a different generalization:
In addition to the simplicial polytopes (related to $A_n$) we obtain as well
hyperoctahedral polytopes (related to $D_n$) for WDVV~solutions in any
dimension, by letting $i{<}j$ in~(\ref{octaroots}) run up to~$n$.\footnote{
Note that our covectors (plus their negatives) form the edges of these
polytopes and not their vertices.}
Such a configuration consists of $n(n{-}1)$ covectors plus their negatives,
but is completely determined again by $n$ of these, for which
$\sfrac12n(n{+}1)$ parameters are needed. Beyond $n{=}3$ the homogeneity
condition~(\ref{flip}) no longer fixes the $f$~coefficients.
The WDVV equation now demands not only that $v_i{\cdot}v_i=v_j{\cdot}v_j$
but also that $\a_{(\pm i\pm j)}{\cdot}\a_{(\pm k\pm l)}=0$
for all indices mutually different.
This is strong enough to enforce $v_i{\cdot}v_j\propto\de_{ij}$,
i.e.~complete regularity for the hyperoctahedron.
What remains for $n{>}3$ is just the $D_n$~root system (up to scale).

Our findings suggest that covector configurations corresponding to
deformations of other roots systems may solve the WDVV equations as well.
For verification, we propose to consider the polytopes associated with the
weight systems of a given Lie algebra, since their edge sets are built from
the root covectors. The idea is then to relax the angles of such polytopes
and analyze the constraints from the homogeneity and WDVV~equations.
The $n$-dimensional hyper-tetrahedra and -octahedra we found emerge simply from
the fundamental and vector representations of $A_n$ and $D_n$, respectively.
Extending this strategy to other representations and Lie algebras
could lead to many more solutions.
\vspace{1cm}

\section{$U{\neq}0$ solutions: 
three-particle systems with $U_{\rm hom}{\equiv}0$}
\noindent
Let us finally try to turn on the other prepotential, $U{\neq}0$, 
in the background of the $F$~solutions already found.
Unfortunately, we have no good strategy to solve (\ref{UF})
unless $U_{\rm hom}{\equiv}0$. 
Hence, in this section let us make the ansatz
\be \label{Uansatz}
U \= -\sum_{\a} g_\a\,\ln|\ax| \ -\ g_R\,\ln R\ ,
\ee
and face the conditions~(\ref{UF2}) (for $n{>}2$) or~(\ref{UF3}) (for $n{=}2$).
In the background of our irreducible root-system solutions,
the Weyl group identifies 
the $f_\a$ and $g_\a$ coefficients for all roots of the same length. Hence, 
besides the $f_{\textrm{L}}$ and $f_{\textrm{S}}$ values in~(\ref{ffamily})
we have couplings $g_{\textrm{L}}$ and $g_{\textrm{S}}$ for a number
$p_{\textrm{L}}$ and $p_{\textrm{S}}$ of long and short positive roots,
respectively.\footnote{
For expliciteness, $p_{\textrm{L}}=\sfrac{n}{2}\,\sfrac{rh^\vee-h}{r-1}$
and $p_{\textrm{S}}=\sfrac{n}{2}\,\sfrac{r(h-h^\vee)}{r-1}$, with the
sum $p=p_{\textrm{L}}+p_{\textrm{S}}=\sfrac{n}{2}h$.}
This simplifies the `sum rule' 
\be \label{neccond}
\sum_{\a} f_\a\,\a{\otimes}\a \= (1{-}f_R)\,\unity
\qquad{\buildrel\textrm{trace}\over\longrightarrow}\qquad
\sum_{\a} \a{\cdot}\a\,f_\a \= n\,(1{-}f_R)
\ee
\be \label{rootsumrule}
\textrm{to}\qquad 2f_{\textrm{L}}\,p_{\textrm{L}}\ +\
\sfrac2r f_{\textrm{S}}\,p_{\textrm{S}}\=n\,(1{-}f_R)
\qquad\buildrel{g_{\textrm{L}},g_{\textrm{S}}\neq0}\over\longrightarrow\qquad
p \= n\,(1{-}f_R)\ .
\ee

We first consider $n{>}2$, hence $g_R{=}0$ and $f_R{=}0$ for $C{\neq}0$.
Since the total number~$p$ of positive roots exceeds~$n$
(except for $A_1^{\oplus n}$), we are forced 
to put either $g_{\textrm{S}}=0$ or $g_{\textrm{L}}=0$.
This fixes all coefficients for $n{\ge}3$ to
\bea \label{case1}
&\textrm{either} \qquad &
g_{\textrm{S}}=0\ ,\quad g_{\textrm{L}}=g 
\qquad\buildrel{(\ref{UF2})(\ref{rootsumrule})}\over\longrightarrow\qquad
f_{\textrm{S}}=\sfrac{r}{2}\sfrac{n-p_{\textrm{L}}}{p_{\textrm{S}}}\ ,\quad
f_{\textrm{L}}=\sfrac12 \\[6pt] \label{case2}
&\textrm{or}\phantom{nnl} \qquad &
g_{\textrm{S}}=g\ ,\quad g_{\textrm{L}}=0
\qquad\buildrel{(\ref{UF2})(\ref{rootsumrule})}\over\longrightarrow\qquad
f_{\textrm{S}}=\sfrac{r}{2}\ ,\quad
f_{\textrm{L}}=\sfrac12\sfrac{n-p_{\textrm{S}}}{p_{\textrm{L}}}\ .
\eea
All simply-laced (ADEH) systems are immediately excluded because they have
$f_\a=\sfrac{1}{h^\vee}$, as is seen in~(\ref{fade}).
In the non-simply-laced (BCFG) one-parameter family (\ref{ffamily}) 
with~(\ref{Fsol}), however, there is always one member which obeys
(\ref{case1}) or~(\ref{case2}) and therefore~(\ref{neccond}).
Furthermore, the trace of~(\ref{UF2}) follows from~(\ref{rootid})
because $g_\a$ and $f_\b$ are constant on $\Phi^+_{\textrm{L}}$
and~$\Phi^+_{\textrm{S}}$.
The same consideration simplifies the expression~(\ref{VB}) for
the bosonic potential at $f_R{=}0{=}g_R$ to
\be \label{VB2}
V_B \= \sum_{\a\in\Phi^+} \frac{v_\a}{(\ax)^2} \ ,\qquad\textrm{where}\qquad
v_\a = \hbar^2\sfrac{f_\a^2}{r_\a^3} \qquad\textrm{or}\qquad
v_\a = \sfrac{1}{r_\a}(g_\a^2{+}\sfrac{\hbar^2}4)
\ee
for any positive root~$\a$ with length${}^2=\frac{2}{r_\a}$,
depending on whether $g_\a$ vanishes or not.
It remains to check the traceless part of~(\ref{UF2}) for the choice
(\ref{case1}) or~(\ref{case2}).
Unfortunately, this is never fulfilled for $n{>}2$, except in the
reducible case of~$A_1^{\oplus n}$.
This failure extends to the deformed root systems,
e.g.~our orthocentric simplex backgrounds.
This rules out $U_{\rm hom}{\equiv}0$ solutions to the flatness
condition for all known irreducible WDVV backgrounds at $n{>}2$.

Therefore, in our search for $C{\neq}0$ solutions~$(F,U)$ 
with $U_{\rm hom}{\equiv}0$, we are forced back to two dimensions, 
i.e.~systems of not more than three particles.
The plethora of $n{=}2$ WDVV solutions~$F$ (parametrized by polygonal chains)
may be cut down by invoking physical arguments. 
If a solution is supposed to describe the relative motion of 
three identical particles, then permuting their coordinates~$x^I$ must be 
equivalent to permuting the covectors (up to sign). 
After separating the center-of-mass coordinate, the planar set~$\{\pm\a\}$
should thus be invariant under the irreducible two-dimensional representation
of~${\cal S}_3$. To visualize the situation, consider the $\R^3$ frame 
rotation by the orthogonal matrix
\be \label{3d2d}
O \= \begin{pmatrix}
\sfrac{1}{\sqrt{2}} &   -\sfrac{1}{\sqrt{2}} & 0 \\[6pt]
\sfrac{1}{\sqrt{6}} & \ph\sfrac{1}{\sqrt{6}} &   -\sfrac{2}{\sqrt{6}} \\[6pt]
\sfrac{1}{\sqrt{3}} & \ph\sfrac{1}{\sqrt{3}} & \ph\sfrac{1}{\sqrt{3}} 
\end{pmatrix}\ :\qquad
e_I\ \buildrel{O}\over\longmapsto\ \sqrt{\frac23}\,\begin{pmatrix}
\cos\f_I \\[6pt] \sin\f_I \\[6pt] \sfrac{1}{\sqrt{2}} \end{pmatrix}
\qquad\textrm{with}\quad \f_I=\frac{2\pi I}{3}-\frac{\pi}{2}\ .
\ee
In the rotated frame, the 3-direction describes the center-of-mass
motion, and the first two entries correspond to the relative-motion plane,
on which the ${\cal S}_3$ representation acts by reflections and
$\sfrac{2\pi}{3}$ rotations.
Reversely, the relative-motion plane is embedded back into the $\R^3$ 
configuration space of the total motion and rotated to the $x^I$~frame via
\be \label{2d3d}
\a_{\text{rel}} \= 
\biggl(\begin{matrix} \cos\f \\[6pt] \sin\f \end{matrix}\biggr)
\quad\hookrightarrow\quad \a_{\text{tot}} \= 
\Biggl(\begin{matrix} \cos\f \\ \sin\f \\ 0 \end{matrix}\Biggr)
\quad\buildrel{O^T}\over\longmapsto\quad
\sqrt{\frac23}\, \Biggl(\begin{matrix}
\sin(\f{+}\frac\pi3) \\ \sin(\f{-}\frac\pi3) \\ -\sin\f \end{matrix}\Biggr) \=
\Biggl(\begin{matrix} \a_1 \\ \a_2 \\ \a_3 \end{matrix}\Biggr)\ ,
\ee
so that the new direction~$(0,0,1)$ becomes the center-of-mass
covector $\rho=\sfrac{1}{\sqrt{3}}(1,1,1)$.
The ${\cal S}_3$ action is generated by
$\f\to\f{+}\sfrac{2\pi}3$ and $\f\to\pi{-}\f$, which produces all permutations 
of the $\a_{\text{tot}}$ entries and hence permutes the~$\{x^I\}$ as required.
The ${\cal S}_3$ orbit of~$\a_{\text{rel}}$ is given by the angle set
\be
\{\,\pm\f\,,\,\pm\f{+}\sfrac{2\pi}3\,,\,\pm\f{-}\sfrac{2\pi}3\,\}
\qquad{\buildrel{\f\ \textrm{special}}\over\longrightarrow}\qquad
\{\,0\,,\,\pm\sfrac{2\pi}3\,\} \quad\textrm{or}\quad
\{\,\pi\,,\,\pm\sfrac\pi3\,\}\ ,
\ee
where the shorter orbits occur for $\f=0$ or $\f=\pi$, modulo~$\sfrac{2\pi}3$.
The upshot is that the two-dimensional covectors must form a 
reflection-symmetric arrangement of $A_2$~systems!

In two dimensions, we take advantage of the radial terms in the structure
equations and turn on all $g$ couplings, which yields (cf.~(\ref{contra}))
\be \label{2dflatness}
f_\a= \sfrac{1}{\a{\cdot}\a}\ \forall\a \ ,\quad
f_R = 1{-}\sfrac{p}{2}\ ,\quad
g_R = \sfrac{p-2}{p}\,C \und\!
\sum_{\a<\b} (g_\a{+}g_\b)\,\frac{\a{\cdot}\b}{\ax\ \bx} = 0
\ee
for some ordering of covectors. 
The bosonic potential~(\ref{VB}) specializes to
\be \label{VB3}
V_B \= \sfrac12\sum_\a \bigl( g_\a^2+\sfrac{\hbar^2}{4}\bigr)
\frac{\a{\cdot}\a}{(\ax)^2} \ +\
\sfrac{p^2{-}4}2\,\bigl(\sfrac{C^2}{p^2}-\sfrac{\hbar^2}4\bigr) \frac1{R^2}
\qquad\textrm{with}\quad R^2=(x^1)^2+(x^2)^2 \ .
\ee
This formula remains correct in the full three-dimensional configuration
space, where one may add the center-of-mass contribution
$V_B^{\text{com}}=\sfrac12 X^{-2}(C_{\text{com}}^2{+}\sfrac{\hbar^2}{4})$.
Please note, however, that $R$ still refers to the relative-motion subspace,
\be
R^2 \quad\hookrightarrow\quad x^T \Bigl( \begin{smallmatrix} 
1 & 0 & 0 \\ 0 & 1 & 0 \\ 0 & 0 & 0 \end{smallmatrix} \Bigr)\, x 
\quad\buildrel{O^T}\over\longmapsto\quad\sfrac13\,x^T\Bigl(\begin{smallmatrix}
\ph 2 &-1 &-1 \\ -1 & \ph 2 &-1 \\ -1 &-1 & \ph 2 \end{smallmatrix} \Bigr)\, x
\ \neq\ {\textstyle\sum_I} (x^I)^2\ .
\ee

Consider now for $\{\a\}$ a collection of $A_2$ systems, each with its own
$g$~value and oriented at a particular angle in the relative-motion plane.
Because each $A_2$ system fulfils the flatness condition by itself, we only
have to compute the `cross terms' in~(\ref{2dflatness}). Introducing the
polar angles $\f_\a$, $\f_\b$ and $\f_x$ of $\a$, $\b$ and $x$, respectively,
the contributions
\be
\frac{\a{\cdot}\b}{\a{\cdot}x\ \b{\cdot}x} \=
\frac{\cos(\phi_\a{-}\phi_\b)}{\cos(\phi_x{-}\phi_\a)\;\cos(\phi_x{-}\phi_\b)}
\=\frac{\tan(\phi_x{-}\phi_\a)-\tan(\phi_x{-}\phi_\b)}{\tan(\phi_\b{-}\phi_\a)}
\ee
to~(\ref{2dflatness}) collapse in telescopic sums, if and only if 
the reflection of any covector on any other one produces again a covector,
and the couplings of mirror-image covectors are identified.
Therefore, the orientations of the various $A_2$ systems must be isotropic, 
i.e.~their collection forms an $I_2(p)$~system with $p=3q$. Ordering the 
positive roots according to their polar angles~$\f_k{=}k\sfrac{\pi}{p}$
with $k=0,1,\ldots,p{-}1$, we get
\be
g_k=g \quad\textrm{for $p$ odd} \qquad\textrm{or}\qquad
g_{2\ell}=g\quad\textrm{and}\quad g_{2\ell+1}=g'\quad\textrm{for $p$ even}\ ,
\ee
so that $\sum_\a g_\a=\sfrac2pC=p\,g$ or $\sfrac{p}2(g{+}g')$, respectively.
Via (\ref{2d3d}) we further obtain
\be \label{3droots}
\frac{\a{\cdot}x}{\sqrt{\a{\cdot}\a}} \quad\to\quad \sqrt{\sfrac23}\,\Bigl(
\sin(k\sfrac{\pi}p{+}\sfrac{\pi}3)\cdot x^1\ +\
\sin(k\sfrac{\pi}p{-}\sfrac{\pi}3)\cdot x^2\ -\
\sin(k\sfrac{\pi}p)\cdot x^3 \Bigr)\ .
\ee
To see a few simple examples, let us give explicit results for $p=3$, 6 and~12.

\medskip\noindent\underline{$A_2$ model.}\\
The minimal model, $p{=}3$, has $f_R{=}{-}\sfrac12$ and $g_R{=}\sfrac13C$
and a single free coupling $g=\sfrac29C$. The radial terms are essential.
In $F$ and~$U$ appear the coordinate combinations
\be \label{a2roots}
\begin{aligned}
\frac{\a{\cdot}x}{\sqrt{\a{\cdot}\a}}\ \in\ \Bigl\{
\sfrac1{\sqrt{2}}(x^1{-}x^2)\,,\,
\sfrac1{\sqrt{2}}(x^1{-}x^3)\,,\,
\sfrac1{\sqrt{2}}(x^2{-}x^3) \Bigr\} \und \\[6pt]
R^2 \ \equiv\ \sfrac13\,x^T\Bigl(\begin{smallmatrix}
\ph 2 &-1 &-1 \\ -1 & \ph 2 &-1 \\ -1 &-1 & \ph 2 \end{smallmatrix} \Bigr)\, x
\= \sfrac13\,\bigl( (x^1{-}x^2)^2+(x^2{-}x^3)^2+(x^3{-}x^1)^2 \bigr)\ ,
\end{aligned}
\ee
so that the bosonic potential becomes
\be \label{A2pot}
V_B \= \bigl(g^2{+}\sfrac{\hbar^2}{4}\bigr)\,\biggl(
\frac{1}{(x^1{-}x^2)^2}+\frac{1}{(x^2{-}x^3)^2}+\frac{1}{(x^3{-}x^1)^2}\biggr)
\ +\ \sfrac58\,\bigl(9g^2{-}\hbar^2\bigr)\,\frac{1}{R^2} \ .
\ee

\medskip\noindent\underline{$G_2$ model.}\\
At $p{=}6$, two $A_2$ systems (with couplings $g$ and $g'$)
are superposed with a relative angle of~$\frac{\pi}{6}$. 
With $f_R{=}{-}2$ and $g_R{=}\sfrac23C$ one has $g{+}g'=\sfrac19C$.
We read off the combinations
\be \label{g2roots}
\begin{aligned} 
\frac{\a{\cdot}x}{\sqrt{\a{\cdot}\a}}\ \ \in\ \ &\Bigl\{
\sfrac{x^1{-}x^2}{\sqrt{2}}\,,\,
\sfrac{2x^1{-}x^2{-}x^3}{\sqrt{6}}\,,\,
\sfrac{x^1{-}x^3}{\sqrt{2}}\,,\,
\sfrac{x^1{+}x^2{-}2x^3}{\sqrt{6}}\,,\,
\sfrac{x^2{-}x^3}{\sqrt{2}}\,,\,
\sfrac{-x^1{+}2x^2{-}x^3}{\sqrt{6}} \Bigr\} \qquad\textrm{and} \\[6pt]
 R^2 &\= \sfrac13\,\bigl( (x^1{-}x^2)^2 + \textrm{cyclic} \bigr) 
\= \sfrac19\,\bigl( (2x^1{-}x^2{-}x^3)^2 + \textrm{cyclic} \bigr) 
\end{aligned}
\ee
and obtain
\be
V_B \= \frac{g^2{+}\sfrac{\hbar^2}{4}}{(x^1{-}x^2)^2}\ +\ 
\frac{3\,({g'}^2{+}\sfrac{\hbar^2}{4})}{(2x^1{-}x^2{-}x^3)^2}\ +\
\textrm{cyclic}\ +\ \frac{36(g{+}g')^2{-}4\hbar^2}{R^2} \ .
\ee

\medskip\noindent\underline{$I_2(12)$ model.}\\
Integrable three-particle models based on $A_2$ and $G_2$ have been
discussed in the literature before.  
Among the infinity of novel models, we take $p{=}12$,
which yields $f_R{=}{-}5$ and $g_R{=}\sfrac56C$, thus $g{+}g'=\sfrac1{36}C$.
In addition to the positive roots of the $G_2$ model (now all `even' 
with coupling~$g$), we have six `odd' roots (with coupling~$g'$),
\be
\frac{\a{\cdot}x}{\sqrt{\a{\cdot}\a}}\bigg|_{\textrm{odd}} \in\ \Bigl\{
\sfrac{\tau x^1{-}x^2{-}\bt x^3}{\sqrt{3}}\,,\,
\sfrac{\tau x^1{-}\bt x^2{-}x^3}{\sqrt{3}}\,,\,
\sfrac{x^1{+}\bt x^2{-}\tau x^3}{\sqrt{3}}\,,\,
\sfrac{\bt x^1{+}x^2{-}\tau x^3}{\sqrt{3}}\,,\,
\sfrac{-\bt x^1{+}\tau x^2{-}x^3}{\sqrt{3}}\,,\,
\sfrac{-x^1{+}\tau x^2{-}\bt x^3}{\sqrt{3}} \Bigr\} 
\ee
where $\tau=\sfrac12(\sqrt{3}{+}1)$ and $\bt=\sfrac12(\sqrt{3}{-}1)$.
The bosonic potential reads
\bea
V_B &=& \frac{g^2{+}\sfrac{\hbar^2}{4}}{(x^1{-}x^2)^2}\ +\
\frac{3\,(g^2{+}\sfrac{\hbar^2}{4})}{(2x^1{-}x^2{-}x^3)^2}\ +\
\frac{\frac32\,({g'}^2{+}\sfrac{\hbar^2}{4})}{(\tau x^1{-}\bt x^2{-}x^3)^2}\ +\
\frac{\frac32\,({g'}^2{+}\sfrac{\hbar^2}{4})}{(\tau x^1{-}x^2{-}\bt x^3)^2}\ +\
\textrm{cyclic} \nonumber \\[6pt] 
&&\ +\ \frac{630(g{+}g')^2{-}\frac{35}{2}\hbar^2}{R^2} \ .
\eea

As has been displayed in~(\ref{a2roots}) and~(\ref{g2roots}), 
for permutation symmetric models the radial coordinate~$R$ may be expressed 
via any triple~$\Gamma$ of roots related by $\sfrac{\pi}{3}$ rotations,
\be
\sum_{\a\in\Gamma}\frac{\a{\otimes}\a}{\a{\cdot}\a} \= \sfrac32\,\unity
\qquad\longrightarrow\qquad
R^2 \= \sfrac23\sum_{\a\in\Gamma}\frac{(\a{\cdot}x)^2}{\a{\cdot}\a}\ ,
\ee
so that, for instance, the radial parts of the prepotentials~(\ref{Fansatz})
may be rewritten as
\be
F_R \= -\sfrac16 f_R\,
\Bigl(\sum_{\a\in\Gamma}\frac{(\a{\cdot}x)^2}{\a{\cdot}\a}\Bigr)\,
\ln\Bigl(\sum_{\a\in\Gamma}\frac{(\a{\cdot}x)^2}{\a{\cdot}\a}\Bigr)\ ,\quad 
U_R \= -\sfrac12 g_R\,
\ln\Bigl(\sum_{\a\in\Gamma}\frac{(\a{\cdot}x)^2}{\a{\cdot}\a}\Bigr)\ .
\ee
The appearance of {\it sums of roots\/} under the logarithm is new.

We further comment that the radial terms for $I_2(3q)$ models with $q$~even
can be eliminated in two ways.
First, choosing $g{+}g'=0$ and the classical limit $\hbar\to0$, one obtains
a conventional model (with covector terms only), but at the expense of
putting~$C{=}0$. 
Second, taking $g'=0$ we can relax the condition $\a{\cdot}\a f_\a{=}1$ for
the odd roots and thus put $f_R{=}0{=}g_R$ in this case, which then yields
$\a{\cdot}\a f_\a=\sfrac{4-p}{p}$ for the odd roots and fixes $g=\sfrac2pC$. 
The bosonic potential in this special situation becomes
\be
V_B \= \sfrac12\,\bigl(g^2+\sfrac{\hbar^2}{4}\bigr)
\sum_{\a\ \textrm{even}} \frac{\a{\cdot}\a}{(\ax)^2}\ +\ 
\sfrac{\hbar^2}{8}\,\sfrac{(4{-}p)^2}{p^2}\,
\sum_{\a\ \textrm{odd}} \frac{\a{\cdot}\a}{(\ax)^2}\ ,
\ee
so in the classical limit only half of the roots remain.
Please note that this result differs from any limit 
of the generic case~(\ref{VB3}).
Of course, the role of even and odd roots may be interchanged.
The results of \cite{wyl} and~\cite{glp2} describe examples of this kind.

The other dihedral groups may also be used to construct three-particle models,
which however lack the permutation symmetry. 
Again we give a couple of prominent examples:

\medskip\noindent\underline{$A_1{\oplus}A_1$ model.}\\
This model is reducible from the outset. From $p{=}2$ it follows that
$f_R{=}0{=}g_R$ so that $g{+}g'=C$. The two orthogonal positive roots
are mapped via~(\ref{3droots}) to
\be
\frac{\a{\cdot}x}{\sqrt{\a{\cdot}\a}}\ \in\ \Bigl\{
\sfrac1{\sqrt{2}}(x^1{-}x^2)\,,\, 
\sfrac1{\sqrt{6}}(x^1{+}x^2{-}2x^3) \Bigr\}\ ,
\ee
and one finds
\be
V_B \= \frac{g^2{+}\sfrac{\hbar^2}{4}}{(x^1{-}x^2)^2}\ +\
\frac{3\,({g'}^2{+}\sfrac{\hbar^2}{4})}{(x^1{+}x^2{-}2x^3)^2}\ .
\ee
Adding the cyclic permutations, one seems to arrive at the $G_2$~model
but cannot produce the (necessary) radial term in this manner.

\medskip\noindent\underline{$BC_2$ model.}\\
The $p{=}4$ case features angles of~$\sfrac{\pi}{4}$. With $f_R{=}{-}1$,
$g_R{=}\sfrac12C$ and $g{+}g'=\sfrac14C$, the one-forms
\be
\frac{\a{\cdot}x}{\sqrt{\a{\cdot}\a}} \in \Bigl\{
\sfrac1{\sqrt{2}}(x^1{-}x^2)\,,\, 
\sfrac1{\sqrt{3}}(\tau x^1{-}\bt x^2{-}x^3)\,,\,
\sfrac1{\sqrt{6}}(x^1{+}x^2{-}2x^3)\,,\,
\sfrac1{\sqrt{3}}(-\bt x^1{+}\tau x^2{-}x^3) \Bigr\}
\ee
enter in
\be
V_B = \frac{g^2{+}\sfrac{\hbar^2}{4}}{(x^1{-}x^2)^2} +
\frac{3\,(g^2{+}\sfrac{\hbar^2}{4})}{(x^1{+}x^2{-}2x^3)^2} +
\frac{\frac32\,({g'}^2{+}\sfrac{\hbar^2}{4})}{(\tau x^1{-}\bt x^2{-}x^3)^2} +
\frac{\frac32\,({g'}^2{+}\sfrac{\hbar^2}{4})}{({-}\bt x^1{+}\tau x^2{-}x^3)^2}+
\frac{6(g{+}g')^2{-}\frac{3}{2}\hbar^2}{R^2} \ .
\ee
Again, this looks like a truncation of the model with $p\to3p$.
Regarding the explicit form of the above expressions, those are unique
only up to rotations around the center-of-mass axis 
$\rho=\sfrac1{\sqrt{3}}(1,1,\ldots,1)$. Our convention has been to
take the first root as $e_1\in\R^2$, which maps to 
$\sfrac1{\sqrt{2}}(e_1{-}e_2)\in\R^3$ under~$O^T$ in~(\ref{2d3d}).

The given examples should suffice to illustrate the general pattern
of dihedral $n{=}2$ solutions with $U_{\rm hom}{\equiv}0$:
The root systems of odd or even order give rise to one- or 
two-parameter three-particle models, which are permutation invariant
only when the order is a multiple of~three. Except for the reducible case
of~$I_2(2)$, the radial contributions are needed; they may disappear only 
when one of the two couplings in the even case vanishes.
\vspace{1cm}

\section{$U{\neq}0$ solutions: three-particle systems in full}
\noindent
Any solution~$U$ (including the trivial $U{\equiv}0$ one) 
for a given $F$~background can be modified by adding to it a homogeneous 
function $\uh$ satisfying $x^i\pa_i\uh=0$ and (\ref{UF}) for $C{=}0$.
As we have seen in the previous section, including this freedom is
in fact mandatory for finding $n{>}2$ solutions in the first place.
In the three-particle case ($n{=}2$), however, we have identified an
infinite series of special solutions, for which we now investigate the
corresponding extension by~$\uh$. In effect, this will add one additional
coupling parameter to the models of the previous section. 

To construct $\uh$ for a rank-two system specified by $\{\a,f_\a,f_R\}$, 
it suffices to solve (\ref{Utrace}) for~$C{=}0$, so that $f_R$ drops out.
As $\uh$ depends only on the ratio $x^2/x^1$ 
we change to polar angles $\phi$ and $\phi_\a$ via
\be
\sfrac{x^2}{x^1} \= \tan\phi \und
\sfrac{\a\cdot x}{\sqrt{\a\cdot\a}} \= R\,\cos(\phi{-}\phi_\a)
\ee
and arrive at
\be \label{Uhom2phi}
\uh''(\phi) - h(\phi)\,\uh'(\phi) \=0 \qquad\textrm{with}\qquad
h(\phi) \= \sum_\a f_\a\,\a{\cdot}\a\,\tan(\phi{-}\phi_\a)\ .
\ee
This is easily integrated (with an integration constant~$\la$) to
\be \label{Uhom2sol}
\uh'(\phi)\=\la\,\prod_\a\bigl[\cos(\phi{-}\phi_\a)\bigr]^{-f_\a\,\a{\cdot}\a}
\ \propto\ R^{2(1-f_R)}\,\prod_\a (\ax)^{-f_\a\,\a{\cdot}\a}
\ee
and blows up on the lines orthogonal to the covectors~$\a$.
Generically, the singularities are $\sim(\ax)^{-1}$.
Only in case some $g_\a$ vanishes, 
the corresponding $f_\a\a{\cdot}\a$ need not equate to one, 
thus $\uh'$ may have a more general singularity structure.
For a dihedral configuration with nonvanishing couplings~$g_\a$
we can go further since 
$\phi_\a=k\sfrac{\pi}{p}$ with $k=0,\ldots,p{-}1$, which yields
\be
h(\phi) = \begin{cases}\!\ph p\,\tan(p\phi) & \textrm{for $p$ odd} \\[4pt]
                       \!-p\,\cot(p\phi) & \textrm{for $p$ even} \end{cases}
\Biggr\}\qquad\longrightarrow\qquad
\uh'= \begin{cases} 
\la\,\bigl[\cos(p\phi)\bigr]^{-1} & \textrm{for $p$ odd} \\[4pt]
\la\,\bigl[\sin(p\phi)\bigr]^{-1} & \textrm{for $p$ even} \end{cases}
\Biggr\}
\ee
and thus (`$\simeq$' means `modulo constant terms')
\be
\uh(\phi) \ \simeq\ \sfrac1p \la\,\ln|\tan(\sfrac{p}2\phi{+}\delta)|
\qquad\textrm{with}\qquad 
\delta\=\Bigl\{\begin{matrix} \sfrac\pi4 & \textrm{for $p$ odd} \\
0 & \textrm{for $p$ even} \end{matrix} \ .
\ee

This may be compared with the particular solution~(\ref{Uansatz}),
\be
U_{\rm{part}} \= -\sum_\a g_\a\,\ln|\a{\cdot}x| - g_R\,\ln R\ \simeq\
-\sum_\a g_\a\,\ln|\cos(\phi{-}\phi_\a)| - C\,\ln R \ ,
\ee
which, in the dihedral case, can be simplified to
(remember that $\sum_\a g_\a{+}g_R=C$)
\be
U_{\rm{part}} \ \simeq\  -C\,\ln R \ -\ \begin{cases} 
g\,\ln|\cos(p\phi)| & \textrm{for $p$ odd} \\[4pt]
g\,\ln|\cos(\sfrac{p}{2}\phi)| + g'\ln|\sin(\sfrac{p}{2}\phi)| 
& \textrm{for $p$ even} \ .
\end{cases}
\ee
Combining $U_{\rm{part}}+\uh=U$ and
lifting to the full configuration space $\R^3\ni(x^I)$, we find
\be \label{pauhom}
\pa_I U \= -\sum_\a g_\a\,\frac{\a_i}{\ax}\ -\ \sfrac{p-2}{p}C\frac{x_i}{R}
\ +\ \la\,\biggl(
\begin{smallmatrix} x^2{-}x^3 \\ x^3{-}x^1 \\ x^1{-}x^2 \end{smallmatrix}
\biggr)\,R^{p-2}\,\prod_\a (\ax)^{-1}\ .
\ee

For the simplest dihedral example, the $A_2$ system, with ($x^{ij}:=x^i{-}x^j$)
\be
F \= -\sfrac14 \bigl[ 
(x^{12})^2\ln|x^{12}|+(x^{23})^2\ln|x^{23}|+(x^{31})^2\ln|x^{31}| \bigr]
\ +\ \sfrac14\,R^2\ln R\ ,
\ee
one gets
\be
\pa_I U \= [x^{12}x^{23}x^{31}]^{-1}\,\biggl(\begin{smallmatrix} 
[\la R-g(x^{31}{-}x^{12})]\,x^{23} \\
[\la R-g(x^{12}{-}x^{23})]\,x^{31} \\
[\la R-g(x^{23}{-}x^{31})]\,x^{12} \end{smallmatrix} \biggr)
\ -\ \sfrac32 g\,R^{-2}\,
\biggl(\begin{smallmatrix} x^1 \\ x^2 \\ x^3 \end{smallmatrix} \biggr)\ ,
\ee
which extends the bosonic potential~(\ref{A2pot}) to
\be
\begin{aligned}
V_B \=\ &\bigl(g^2{+}\sfrac23\la^2{+}\sfrac{\hbar^2}{4}\bigr)\,\biggl(
\frac{1}{(x^{12})^2}+\frac{1}{(x^{23})^2}+\frac{1}{(x^{31})^2}\biggr) \\[4pt]
&+\ \sfrac58\,\bigl(9g^2{-}\hbar^2\bigr)\,\frac{1}{R^2}\ +\
\la\,g\,R\,\frac{(x^{12}{-}x^{23})(x^{23}{-}x^{31})(x^{31}{-}x^{12})}
{(x^{12}x^{23}x^{31})^2}\ .
\end{aligned}
\ee
\vspace{1cm}

\section{Conclusion}
\noindent
In this paper we systematically constructed conformal $(n{+}1)$-particle 
quantum mechanics in one space dimension with $\cN{=}4$ supersymmetry, 
i.e.~$su(1,1|2)$ invariance, and a central charge~$C$. To begin with, 
the closure of the superalgebra produced a set of `structure equations'
(\ref{w1}) and~(\ref{w2}) for two scalar prepotentials $U$ and~$F$,
which determine the potential schematically as 
$V=\sfrac12U'U'+\sfrac{\hbar^2}8F^{\prime\prime\prime}F^{\prime\prime\prime}$
plus fermionic terms. The structure equations consist of homogeneity
conditions depending on~$C$, a (generalized) WDVV~equation (for $F$ alone) 
and a `flatness condition' (for $U$ in the $F$~background).

Separating the center-of-mass degree of freedom reduces the configuration
space from $\R^{n+1}$ to~$\R^n$ for the relative motion.
The ansatz~(\ref{Fansatz}) for the many-body functions $U$ and~$F$ turned 
the structure equations into a decomposition of the identity~(\ref{hcond}) and 
nonlinear algebraic relations (\ref{FF}) and~(\ref{UF2}), for a set $\{\a\}$ 
of covectors in~$\R^n$ and real coupling coefficients $g_\a$ (for~$U$) and 
$f_\a$ (for~$F$). The homogeneous part of~$U'$ is governed by a 
linear differential equation~(\ref{UF}) (with $C{=}0$) of Fuchsian type. 
The case of three particles is special, because the WDVV~equation is empty
and so anything goes for~$F$, but the flatness condition for~$U$ is still
nontrivial.

To find the prepotential~$F$ it suffices to solve the
WDVV~equation~(\ref{FF}). It is known that the roots of any finite 
reflection group provide a solution~\cite{magra,vese}, each giving rise 
to an interacting quantum mechanics model with~$U{\equiv}0$ and thus $C{=}0$. 
Besides rederiving this result in a new fashion, we were able to generalize 
it in two ways:
First, the $A_n$ root system may be deformed to a system of edges for
a general orthocentric $n$-simplex, yielding a nontrivial $n$-parameter
family of WDVV~solutions which might agree with one found in~\cite{chaves}.
Second, the relative weights for the long and the short roots contributing
to~$F$ are undetermined even in sign, so that the $BCF$-type solutions form
one-parameter families. 

For a nonzero central charge, in any given $F$~background one must turn on
the prepotential~$U$ by solving~(\ref{UF}). Within our ansatz~(\ref{Fansatz}),
this requires finding a suitable homogeneous part~$\uh$ -- an unsolved task.
Only if in appropriate coordinates the system decomposes into subsystems
not larger than rank two, then $\uh$ is not needed but can easily be found.
Thus for the special case of three particles, i.e.~$n{=}2$, the situation is 
simpler: the flatness condition~(\ref{trace}) then permits the novel `radial
terms' which provided the necessary flexibility in our ansatz~(\ref{Fansatz}). 
Again the covectors were forced into a root system, which as of rank two 
must be dihedral. We explicitly constructed the full prepotentials 
(including~$\uh$) for the new infinite dihedral series and 
displayed several examples lifted back to the original configuration
space~$\R^3\ni(x^1,x^2,x^3)$. When the dihedral group and the central charge
are fixed, the model depends on one or two tunable coupling parameters
depending on the group order~$p$ being odd or even. Permutation symmetry
requires $p$ to be a multiple of~3.
The previously found models~\cite{wyl,glp2} turned out to be either 
decomposable or peculiar special cases of our dihedral systems, 
for which the `radial terms' could by omitted. 
To summarize, we have classified all one-dimensional $\cN{=}4$ superconformal 
quantum three-particle models based on covectors.

It remains an open problem to construct any irreducible $U{\neq}0$ solutions 
with more than three particles and to find all $U{\equiv}0$ solutions,
i.e.~the complete moduli space of the WDVV~equation.
To complement recent progress in mathematics on this issue~\cite{feives2},
we would like to propose another strategy towards this goal:
take any simple Lie algebra, select one of its irreducible representations
and form the convex hull of its weight system. The edges of this polytope
reproduce the roots, with certain multiplicities. Now consider a deformation
of this polytope. Generically, the degeneracy of the edge orientations
will be lifted, but the deformed collection of covectors still satisfies
the incidence relation of the polytope. We suggest to test the WDVV~equation
on such configurations, generalizing the method successful for the fundamental
$A_n$~representation. We are confident that this is feasible and will lead
to further beautiful mathematical structures.
\vspace{0.5cm}

\section*{Acknowledgments}
\noindent
O.L. thanks N.~Beisert and W.~Lerche for useful discussions.
A.G. is grateful to the Institut f\"ur Theoretische Physik at the
Leibniz Universit\"at Hannover for hospitality and to DAAD for support.
The research was supported by RF Presidential grant MD-2590.2008.2,
NS-2553.2008.2, DFG grant 436 RUS 113/669/0-3 and the Dynasty Foundation.
\vspace{0.5cm}

\section*{Note added}
\noindent
After a first version of this work had appeared on the arXiv, several aspects
discussed here have been developed further in the three related
papers \cite{bks}--\cite{klp}.

\end{document}